\documentclass[12pt,a4paper,oneside]{article}
\usepackage[utf8]{inputenc}
 \usepackage{times}

\usepackage{amssymb,amsthm,amsmath}
\usepackage{setspace}
\usepackage{a4wide}

\usepackage[misc]{ifsym} 

\usepackage[backend=biber,style=nature,citestyle=numeric, sorting=none]{biblatex}
\usepackage{graphicx}
\usepackage[font=sf]{floatrow}

\usepackage{paralist}  
\renewcommand{\figurename}{Fig.}
\renewcommand{\tablename}{Tab.}

\usepackage{longtable}
\usepackage{booktabs}
\usepackage{makecell, cellspace}
\usepackage{array}
\newcolumntype{L}[1]{>{\raggedright\let\newline\\\arraybackslash\hspace{0pt}}m{#1}}
\newcolumntype{C}[1]{>{\centering\let\newline\\\arraybackslash\hspace{0pt}}m{#1}}
\newcolumntype{R}[1]{>{\raggedleft\let\newline\\\arraybackslash\hspace{0pt}}m{#1}}

\makeatletter
\let\@fnsymbol\@arabic

\usepackage{lineno}

\title{The Science of Startups: The Impact of Founder Personalities on Company Success}

\author{
    Paul X. McCarthy\thanks{The Data Science Institute, University of Technology Sydney, NSW, Australia. \textsuperscript{\Letter}paul@onlinegravity.com}\,\,\,\textsuperscript{,}\thanks{School of Computer Science and Engineering, UNSW Sydney, NSW, Australia.}\,\,\textsuperscript{,}\,\textsuperscript{\Letter}\,, 
    Xian Gong\thanks{Faculty of Engineering and Information Technology, University of Technology Sydney, Australia}\,\,,
    Fabian Stephany\thanks{Oxford Internet Institute, University of Oxford, Oxford, UK. \textsuperscript{\Letter}fabian.stephany@oii.ox.ac.uk}\,\,\,\textsuperscript{,}\thanks{DWG Datenwissenschaftliche Gesellschaft Berlin, Germany.\textsuperscript{\Letter}fabian.braesemann@oii.ox.ac.uk}\,\,\textsuperscript{,}\,\textsuperscript{\Letter}\,, \\
    Fabian Braesemann\footnotemark[4]\,\,\,\textsuperscript{,}\footnotemark[5]\,\,\textsuperscript{,}\,\textsuperscript{\Letter}\,
    Marian-Andrei Rizoiu\footnotemark[3]\,\,,
    Margaret L. Kern\thanks{Melbourne Graduate School of Education, The University of Melbourne, Parkville, VIC, Australia.}\\
    }

\begin{document}

\maketitle
\doublespacing

\begin{abstract}
    Startup companies solve many of today’s most complex and challenging scientific, technical and social problems, such as the decarbonisation of the economy\cite{goldstein2020patenting}, air pollution\cite{lewis2016validate}, and the development of novel life-saving vaccines\cite{mulligan2020phase}. Startups are a vital source of social, scientific and economic innovation, yet the most innovative are also the least likely to survive\cite{hyytinen2015does}. The probability of success of startups has been shown to relate to several firm-level factors such as industry, location and the economy of the day\cite{zbikowski2021machine}. Still, attention has increasingly considered internal factors relating to the firm’s founding team, including their previous experiences and failures\cite{yin2019quantifying}, their centrality in a global network of other founders and investors\cite{bonaventura2020predicting} as well as the team’s size\cite{klotz2014new}. The effects of founders’ personalities on the success of new ventures are mainly unknown. Here we show that founder personality traits are a significant feature of a firm’s ultimate success. We draw upon detailed data about the success of a large-scale global sample of startups (n=26,781).  We found that the Big 5 personality traits of startup founders across 30 dimensions significantly differed from that of the population at large. We can train a classifier to distinguish founders from employees with 82.5\% accuracy. Key personality facets that distinguish successful entrepreneurs include a preference for variety, novelty and starting new things (openness to adventure), like being the centre of attention (lower levels of modesty) and being exuberant (higher activity levels). However, we do not find one “Founder-type” personality; instead, six different personality types appear, with startups founded by a “Hipster, Hacker and Hustler” being twice as likely to succeed. Our results also demonstrate the benefits of larger, personality-diverse teams in startups, which has the potential to be extended through further research into other team settings within business, government and research. 
\end{abstract}

\section*{Background}

The success of startups is vital to economic growth and renewal, with a small number of young, high-growth firms creating a disproportionately large share of all new net jobs\cite{henrekson2010gazelles}. Startups create jobs and drive economic growth, and they are also an essential vehicle for solving some of society's most pressing challenges. 

As a poignant example, six centuries ago, the German city of Mainz was abuzz as the birthplace of the world's first moveable-type press created by Johannes Gutenberg. However, in the early part of this century, it faced several economic challenges, including rising unemployment and a significant and growing municipal debt. Then in 2008, two Turkish immigrants formed the company BioNTech in Mainz with another university research colleague. Together they pioneered new mRNA-based technologies. In 2020, BioNTech partnered with US pharmaceutical giant Pfizer to create one of only a handful of vaccines worldwide for Covid-19, saving an estimated six million lives\cite{economist2022jul}.  The economic benefit to Europe and, in particular, the German city where the vaccine was developed has been significant, with windfall tax receipts to the government clearing Mainz’s €1.3bn debt and enabling tax rates to be reduced, attracting other businesses to the region as well as inspiring a whole new generation of startups\cite{oltermann2021pfizer}. 

While stories such as the success of BioNTech are often retold and remembered, their success is the exception rather than the rule. The overwhelming majority of startups ultimately fail. One study of 775 startups in Canada that successfully attracted external investment found only 35\% were still operating seven years later\cite{grant2019survival}. An industry “autopsy” into 101 tech startup failures found 23\% were due to not having the right team — the number three cause of failure ahead of running out of cash or not having a product that meets the market need\cite{reason2019cb}.

\subsection*{Introduction}
In this project, we aimed to understand whether certain combinations of founder personalities are related to startup success, defined as when the firm has been acquired, acquired another firm or is listed on a public stock exchange. The project provides a large-scale quantitative perspective on the colloquial “Hacker, Hustler, Hipster”\cite{ellwood2012dream} dream team that is envisaged to form the optimal combination of personalities to accomplish business success. For the quantitative analysis, we draw on a previously published methodology\cite{kern2019social}, which matched people to their ideal jobs based on social media-predicted personality traits.

Here, we applied the same methodology to another set of Twitter users: founders and executives with a Crunchbase profile. Crunchbase is the world’s largest directory on startups. It provides information about more than 1 million companies, primarily focused on funding and investors. A company's Crunchbase profile can be considered a digital business card of an early-stage venture. As such, the founding teams tend to provide information about themselves, including their educational background or a link to their Twitter account. Again, as with Twitter, all information on Crunchbase is publicly available. 

In this project, we inferred the personality profiles of the founding teams of early-stage ventures using the methodology described from their publicly available Twitter profiles. Then, we correlated this information to funding from Crunchbase to determine whether particular combinations of personality traits correspond to the success of early-stage ventures.

\subsection*{What makes for a successful startup?}
Venture capitalists and other investors, especially in early-stage unproven startup companies, each have their perspective on the key factors that make for likely success. Three different schools of thought can mostly characterise these different perspectives:

\begin{description}
  \item[\textit{Supply-side or product investors}:] those who prioritise investing in firms they consider to have novel and superior products and services, investing in companies with intellectual property such as patents and trademarks.
  \item[\textit{Demand-side or market-based investors}:] those who prioritise investing in areas of highest market interest such as in hot areas of technology like \emph{quantum computing} or recurrent or emerging large-scale social and economic challenges such as \emph{decarbonisation of the economy}.
  \item[\textit{Talent investors}:] those who prioritise the foundation team above the startup’s initial products or what industry or problem it is looking to address. 

\end{description}

Getting to the point at which the startup has demonstrated the market is willing to use and pay for its novel products and services regularly, known as \emph{product-market} fit, is seen as a vital milestone for investors and founders alike, and is often a conditional trigger for additional rounds of investment. 

Much focus in recent years has been on reconciling the first two of these investor perspectives to achieve \emph{product-market} fit as quickly and with the least possible capital invested in creating a minimum viable product. 

However, investors who adopt the third perspective and prioritise talent recognise that a good team can overcome many challenges in the lead-up to \emph{product-market} fit. And while the initial products of a startup may or may not work, a successful and well-functioning team has the potential to pivot to new markets and new products, even if the initial ones prove untenable. 

Some of today’s most prominent startup success stories, such as \emph{Twitter}, were not the startups’ first idea for a product or service but the result of trying several other things that failed. This story is common in product innovation, with many well-known consumer products emerging from previous “failures”. For example, the renowned engineering lubricant WD-40 is so named as the result of the 40th attempt to create the formula, and 3M’s Post-It notes were a product made from a “failed” adhesive project.

In this article, we analyse a variety of \emph{firm-level}, \emph{founder-level} and \emph{founder-team-level} determinants of the success of startups, which are by their very nature experimental, high risk and likely to fail. 

\begin{figure}
  \includegraphics[width=0.7\textwidth]{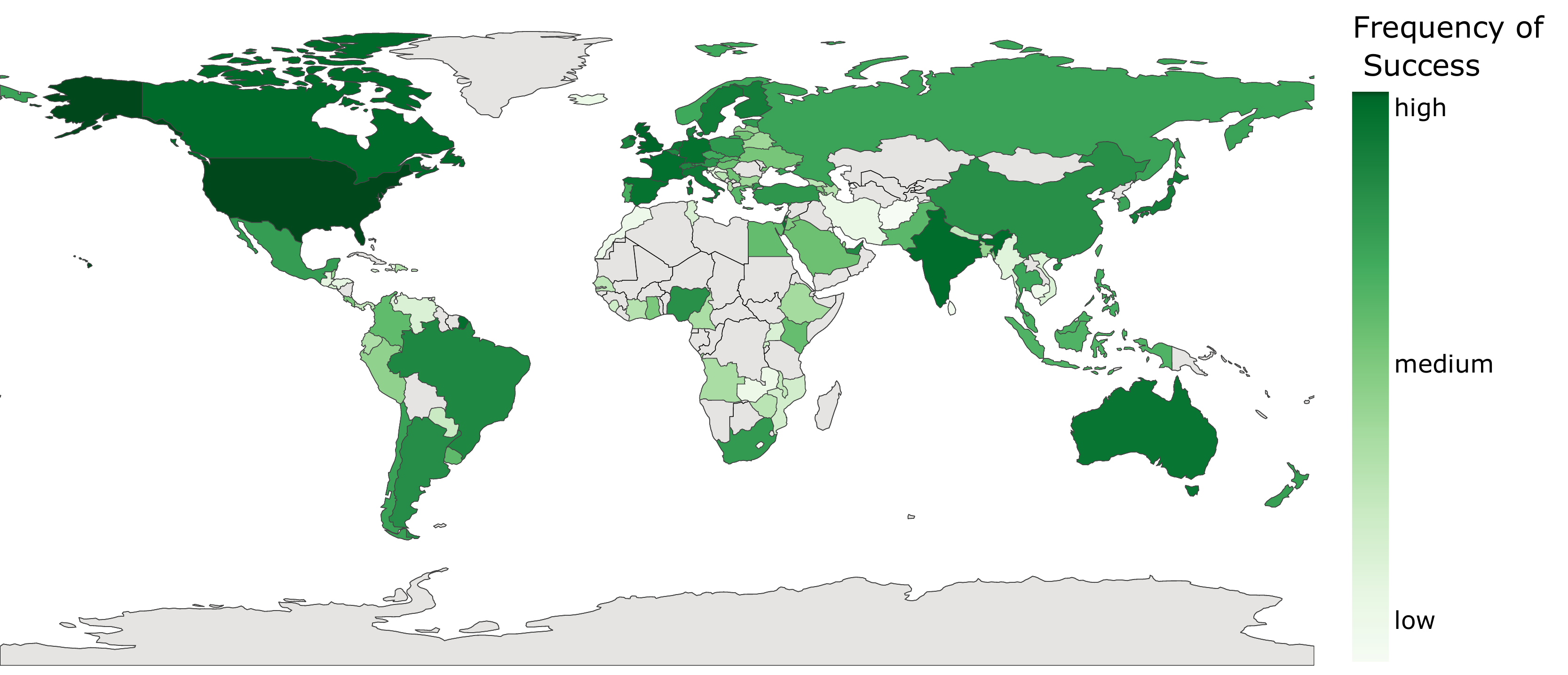}\label{fig:Figure1a}\\[3ex]
  \includegraphics[width=0.7\textwidth]{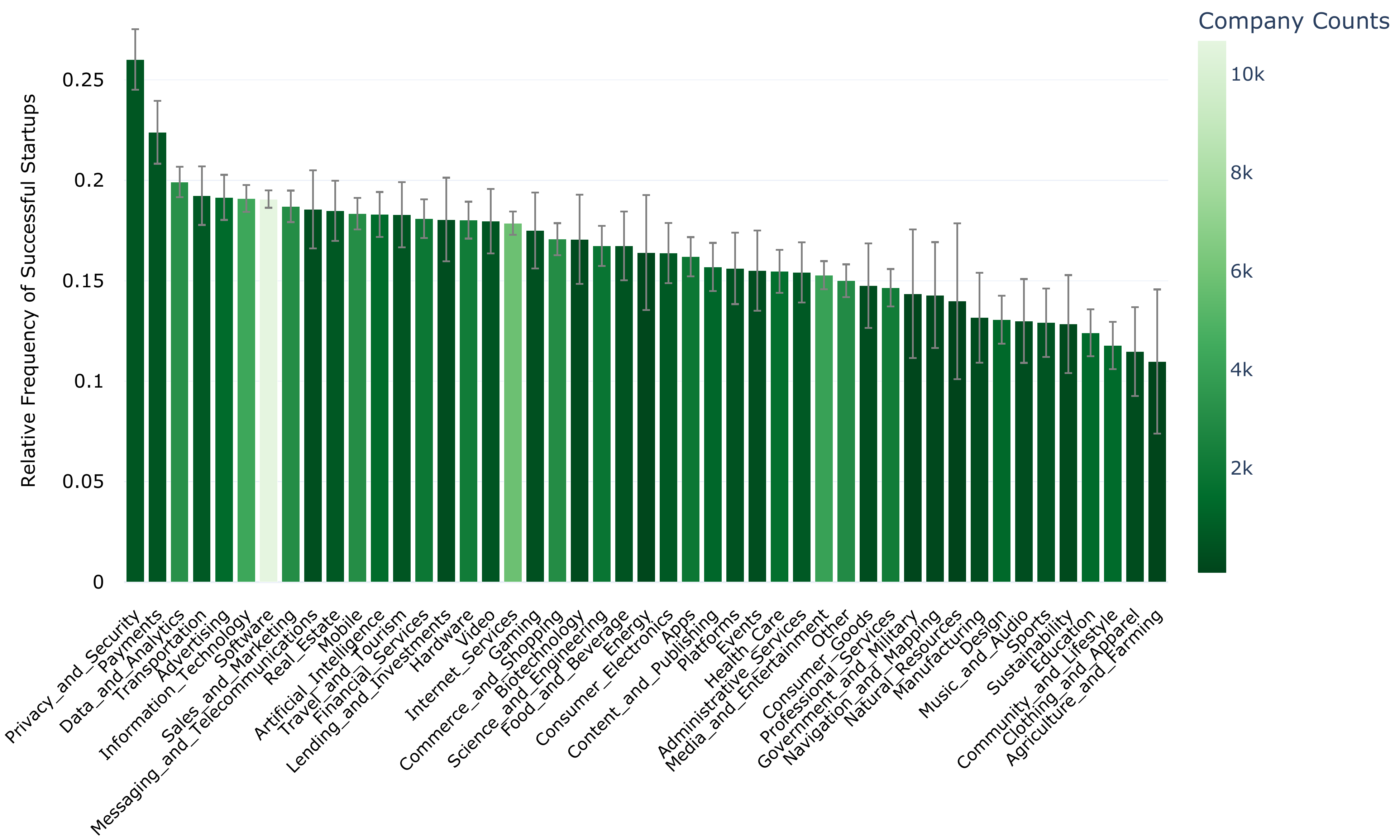}\label{fig:Figure1b}\\[3ex]
  \includegraphics[width=0.7\textwidth]{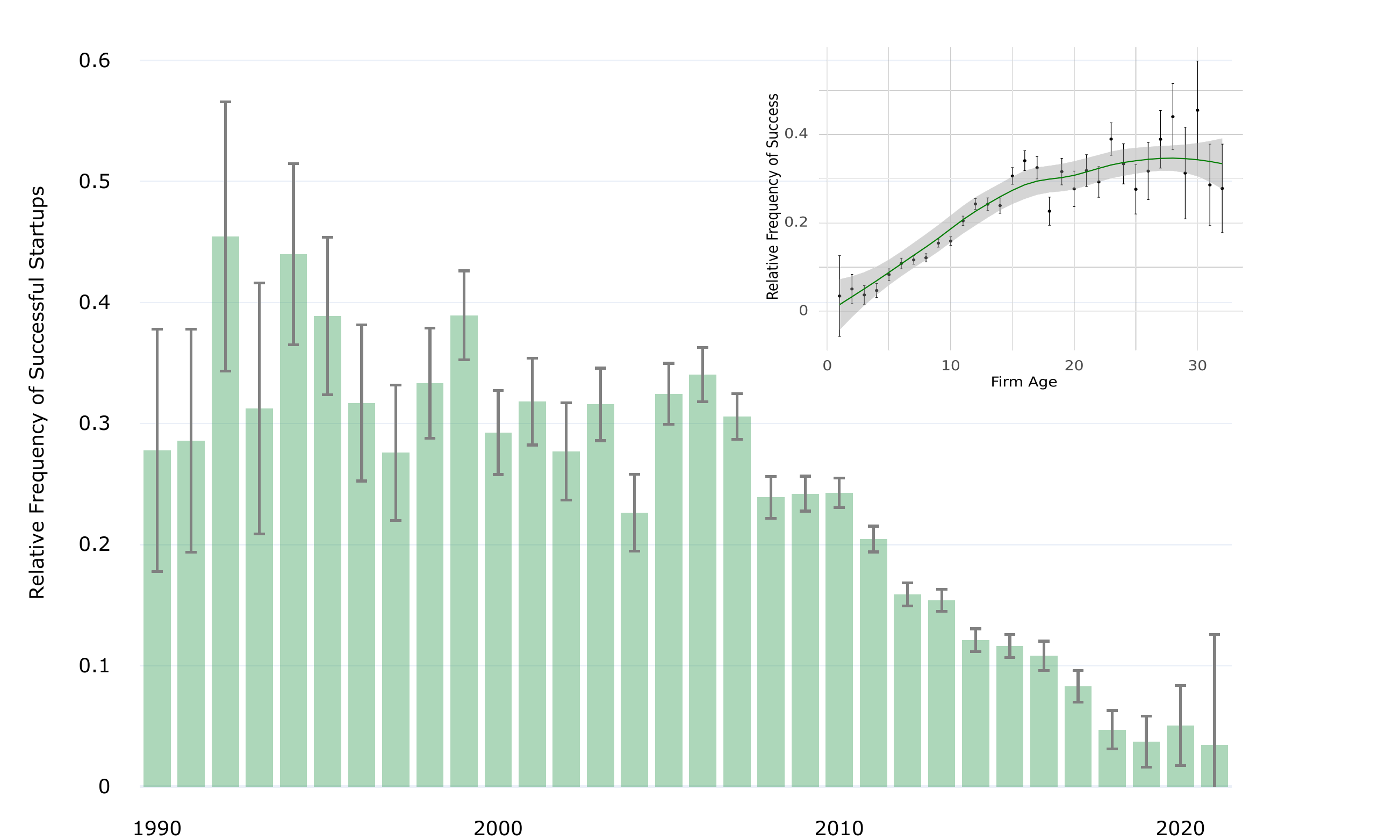}\label{fig:Figure1c}
  \caption{\textbf{$|$ Firm-Level Factors of Startup Success.} \textbf{a}, On a country level, chances for success are highest in the US, Japan, West Europe, and Scandinavian countries. \textbf{b}, Firms from the payment and software industries have high chances of success. \textbf{c}, Chances of success are positively related to a firm's maturity, with firms that are seven years or older having higher chances of success.}\label{fig:Figure1}
\end{figure}

Firstly, we examine a range of \emph{firm-level} determinants of startup success, including \emph{location} (Fig.~\ref{fig:Figure1a}), \emph{industry} (Fig.~\ref{fig:Figure1b}) and \emph{age of startup} (Fig.~\ref{fig:Figure1c}) to explore to what extent these factors are associated with success. Then building on our previous occupation-personality fit research\cite{mccarthy2022occupation}, we use a large collection of public data on startup companies from Crunchbase to examine the detailed personality profiles of founders. Finally, in a series of experiments with large-scale samples, we explore three fundamental questions:

\begin{enumerate}
  \item What, if any, personality features distinguish them as entrepreneurs? And if so, what types of personality combinations exist among startup entrepreneurs?
  \item Does the personality of its founders play a role in a startup’s success when accounting for other external factors known to influence it, such as location, industry and company age? 
  \item Does the combination of founders and their personalities play a role in startup success, and is there any evidence to support the commonly held view in the venture capital investment community that startups require three types of founders: a Hacker, a Hustler and a Hipster? 

\end{enumerate}

\subsection*{The rise of the hipster in startups}
Clear functional roles have evolved in established industries such as film and television, construction and advertising. 

In advertising, there is a long-established functional distinction between the categorical roles of \emph{creatives} (people who devise the words, images and music for advertisements, including copywriters and creative directors), \emph{suits} (client-facing account managers and sales executives) and \emph{quants} (strategy and planning roles associated with audience measurement and the buying and placements of advertisements across different media). 

The necessary tension, especially between \emph{suits} and \emph{creatives} in advertising, is well understood, as “there is an enduring oppositional culture between the ‘\emph{creatives}’ and the ‘\emph{suits}’ within agencies. From the point of view of the ‘\emph{creatives}’, the lifeblood of the agency is considered to lie in the creative team with the other functions either considered inferior or unavoidable evils”\cite{pratt2006advertising}.

In technology, the categorical roles of Hackers (skilful computer programmers and developers) and Hustlers (entrepreneurial leaders able to win over customers and investors to new products and ideas) have been around for decades, with similar oppositional tension. For example, when Steve Jobs announced he would take medical leave from Apple in January 2009, Mat “Wilto” Marquis described him as a hacker and a hustler in a well-wishing tweet.

However, the first use of Hacker and Hustler in conjunction with Hipster in the context of the putative startup founder dream was coined by influential venture capitalist Elias Bizannes in 2011. It was then popularised in 2012 by an address at the influential technology conference South by Southwest by Rei Inamoto and in a subsequent \emph{Forbes} article “\emph{The Dream Team: Hipster, Hacker, and Hustler}”\cite{ellwood2012dream}.

Hipster is a broad term used to describe members of an urban subculture in many cities in the US and other countries who are design conscious and favour non-mainstream fashions, trendy foods and alternative music. Bizannes co-opted the term to reflect what he perceived was the increasing need for successful startups to have a founder with design-savvy, aesthetic imagination and insider knowledge (Hipster) in addition to the traditional roles of someone good at selling things (Hustler) and creating technology products (Hacker). 

\subsection*{Founders are not like most other people}
As a first step, we explore whether the personalities of successful startup founders are measurably different from those of people in other occupations. 

While recent research has demonstrated that many employees in the same occupations share similar personality traits\cite{kern2019social}, being a startup founder is not a conventional job. So while we now have maps of the personality signatures of many jobs, startup founders' personality signatures have yet to be identified. 

Employing established methods\cite{schwartz2013personality, plank2015personality, arnoux201725}, we inferred the personality traits across 30 dimensions (Big 5 facets) of a large global sample (n=4.4k) of successful startup founders. The successful startup founders cohort was created from a subset of self-identified founders from the global startup industry directory Crunchbase, who are also active on the social media platform Twitter and have a record of successfully attracting external venture capital investment.  Success in a startup is typically staged and can appear in different forms and times. For example, a startup may be seen to be successful when it finds a clear solution to a widely recognised problem, such as developing a successful vaccine. On the other hand, it could be achieving some measure of commercial success, such as rapidly accelerating sales or becoming profitable or at least cash positive. Or it could be reaching an exit for foundation investors via a trade sale, acquisition or listing of its shares for sale on a public stock exchange via an Initial Public Offering (IPO). However, one commonly agreed measure of success is the attraction of external investment by venture capitalists.

Many startup founders wear multiple hats. In addition to being startup founders or cofounders, they often perform functional roles (and sometimes hold the titles of conventional C-Suite leaders) such as CEO, CTO or CFO. Some also have full-time or part-time jobs as engineers, managers or consultants in other companies unrelated to their startup while in the early stages of developing their fledgling business.

While not all CEOs are founders (and indeed, most are not), some are also CEOs. Founders much more commonly hold some occupations like CEOs than others, and other job types are rarely held by founders. We use this overlap between startup founders holding conventional roles to create a complementary sample of \emph{successful employees} unlikely to be founders. 

To begin, we leveraged data from previous occupation-fit research on the personality traits of successful employees in 624 different occupations across various industries. 

Then, we developed an \emph{Entrepreneurial Occupational Index (EOI; see Extended Data Fig.~\ref{fig:ExtendedFig21})} based on LinkedIn data that looks at the percentage of people currently employed in that role worldwide and who also hold or have previously held the position of founder or co-founder.  We found EOI values for each of the 624 Occupations we have personality profiles for. We ranked the occupations from most entrepreneurial (public speaker 21.11\%, chief technology officer 20.75\%, and creative director 19.33\%) to least entrepreneurial (cashier 0.02\%, palaeontologist 0.00\%, furniture removalist 0.00\%, aged carer 0.00\%, bacteriologist 0.00\%).

We then created a list of low EOI occupations (n=112), each of which had less than 0.5\% of whom also held the titles founder or co-founders in their \emph{LinkedIn} Profile. People in these roles may still be founders and co-founders, but it is unlikely that they are. Any individual in even the most entrepreneurial of these 112 occupations (internal auditor) is still five times less likely also to be a founder or co-founder than the global average (2.5\%) across all 624 occupations. From our previous study, we randomly selected a sample of \emph{Successful Employees} (n=6k) for whom we have inferred personality data and who are unlikely to be entrepreneurs as they are drawn from the 112 low EOI occupations.  

Using the two samples together: \emph{Successful Entrepreneurs} and \emph{Successful Employees} (unlikely to be founders), we trained and tested a machine learning random forest classifier to distinguish and classify entrepreneurs from employees and vice-versa using inferred personality vectors alone. As a result, we found we could correctly predict \emph{Entrepreneurs} with 77\% accuracy and \emph{Employees} with 88\% accuracy (Fig.~\ref{fig:Figure2a}). Thus, based on personality information alone, we correctly predict all unseen new samples with 82.5\% accuracy (See Extended Data Fig.~\ref{fig:ExtendedFig1} for details on modelling and prediction accuracy.).

\subsubsection*{Adventurousness — the key feature} 
We explored in greater detail which personality features are the most important in distinguishing successful entrepreneurs from successful employees and found that the subdomain or facet of \emph{Adventurousness} within the Big 5 Domain of Openness was both significant and had the largest effect size.  The facet of \emph{Modesty} within the Big 5 Domain of Agreeableness and \emph{Activity Level} within the Big 5 Domain of Extraversion was the subsequent most considerable effect (Fig.~\ref{fig:Figure2b}). All thirty dimensions of the Big 5 facet were found to be significantly different in their distribution, with ten features having large effect sizes. (See Extended Data Table~\ref{tab:ExtendedTable1} for more details of Cohen’s D analysis with a complete list of features and their effect sizes and Extended Data Fig.~\ref{fig:ExtendedFig2} for Big5 personality facets of Employees and Entrepreneurs visualised as a heatmap and dendrogram.)

This is important because, to our knowledge, this is the first study to show differences between employees and entrepreneurs at the facet level of the Big 5 personality domains and the largest-scale study (n=10.4k) of any kind in this field. 

In our sample, \emph{Successful Entrepreneurs} were defined as founders or co-founders of companies who have attracted over USD \$100k+ in investments from venture capitalists. This is consistent with previous research that found higher values in the personality trait Openness significantly predict VC financing even after accounting for observable founder and firm characteristics\cite{chapman2021founder} and the key Big 5 Domain that distinguishes entrepreneurs from non-entrepreneurs\cite{antoncic2015big}. 

Adventurousness in the Big 5 framework is defined as the preference for variety, novelty and starting new things - which are consistent with the role of a startup founder whose role, especially in the early life of the company, is to explore things that do not scale easily\cite{graham2013things} and is about developing and testing new products, services and business models with the market.  

\subsection*{Six types of startup founders}
Once we understood that startup founders have distinctive personality features that are different from regular employees, we explored whether there are distinct types of personalities among startup founders. 

First, we examined whether there is evidence to show that startup founders naturally cluster according to their personality features using a Hopkins test. We discovered clear clustering tendencies in the data compared with other renowned reference data sets known to have clusters.  Specifically, we found that founders' personalities have higher clustering tendency scores than that of two well-known scientific data sets with known in-built clustering: Edgar Anderson's classic detailed measurements of three species of Irises\cite{anderson1935irises} and the more recent size measurements for three species of Pygoscelis penguins that breed on islands throughout the Palmer Archipelago\cite{horst2022palmer} (see Extended Data Fig.~\ref{fig:ExtendedFig3}).

Then, once we established the founder data clusters, we used agglomerative hierarchical clustering; a “bottom-up” clustering technique that initially treats each observation as an individual cluster and then merges them to create a hierarchy of possible cluster schemes with differing numbers of groups (See Extended Data Fig.~\ref{fig:ExtendedFig4}).

And lastly, we identified the optimum number of clusters based on the outcome of four different clustering performance measurements: Davies-Bouldin Index, Silhouette coefficients, Calinski-Harabas Index and Dunn Index.  We found that the optimum number of clusters of startup founders based on their personality features is six (labelled \#0 through to \#5).

\subsubsection*{Personality footprints of founders}
To better understand the unique personality characteristics of each of the six different clusters of founders and co-founders we:

\begin{enumerate}
  \item \textbf{Analysed the personality footprints of each cluster.}  We examined the distinctive personality traits of each group and identified which clusters were home to the maximums in each of the 30 personality facets (See summary in Table~\ref{tab:Table1}) and also created a heat map revealing the complete personality footprint of each of the six types (Fig.~\ref{fig:Figure2d}).
  \item \textbf{Matched the occupation closest to the centre of each cluster} using the personality-occupation matrix from our previous research in two separate studies based on 128,279 people in 3,513 professions using ten dimensions\cite{kern2019social} and a second more recent study based on 99,897 people in 624 occupations using 30 personality dimensions\cite{mccarthy2022occupation}. 
  \item \textbf{Identified which of the eight occupation-tribes from previous research\cite{mccarthy2022occupation} each founder or co-founder belonged to.}  Leveraging previous research, we then looked at the distribution of tribe membership of each founder within each cluster.  

\end{enumerate}

\subsubsection*{Founders within the personality-occupation landscape}
To better understand the context of different founder types, we positioned each of the footprints of each of the six types of founders within an occupation-personality matrix (n=624 jobs) established from previous research\cite{mccarthy2022occupation}. Prior research showed that “each job has its own personality” using a substantial sample of employees (n=99k) across various jobs. Furthermore, we found that the occupations themselves clustered into eight different groups—which we refer to as \emph{occupation tribes} — based on their personality alone.  The key personality attributes of each of these tribes from this prior research is reproduced in Extended Data Fig.~\ref{fig:ExtendedFig22}.

\begingroup
\footnotesize
\singlespacing
\begin{longtable}{ C{2.5cm}  C{5.5cm}  C{4.5cm}  C{3cm} }
\caption{$|$ Typology of Founders by Personality
    \textbf{$|$ Typology of Founders by Personality.} Six different types of founders are revealed by clustering founders (n=32k) by their Big 5 personality facets. Each type\,---\,Fighter, Operator, Accomplisher, Leader, Engineer and Developer (FOALED)\,---\,has its distinctive personality footprint, but three are equivalent to variations of Hackers (Fighters, Operators and Developers), two are variations of Hustlers (Leaders and Accomplishers) and one can be characterised as equivalent to a Hipster (Engineer).}\\
    \toprule
    \makecell{\textbf{Founder Type} \\ Clustered by \\Personality} & \makecell{\textbf{Distinctive Personality Traits} \\ Personality traits of founders \\in  this cluster (Big 5 facets)}  & \makecell{\textbf{Closest Occupation } \\ Occupation maps \\ (Repec\cite{mccarthy2022occupation} and PNAS\cite{kern2019social})} & \makecell{\textbf{3H Typology} \\ Hipster / \\ Hacker / Hustler }\\
    \hline
    \endhead
   Leaders (\#2) & Highest in openness in the facets of artistic interests and emotionality also highest in agreeableness in facets of altruism and sympathy. & Executive Director, Medical Director & Hustler (Pure)\\
  \hline
   Accomplisher (\#0) & Highly extraverted (all facets) and Conscientious (five facets) & Chief Information Officer, Export Manager & Hustler (Technology Focus) \\
   \hline
   Operator (\#4) & Highest in conscientiousness in the facet of orderliness and high agreeableness in the facet of humility for founders in this cluster. & Bicycle Mechanic, Mechanic and Service Manager. & Hacker (Operations focus) \\
    \hline
   Developer (\#3) & “Middle child” cluster — no facets are maximums or minimums, but it shares characteristics similar to fighters but higher in extraversion. & Application Developer and related technology roles such as Business Systems Analyst and Product Manager. & Hacker (Product focus)  \\
    \hline
   Fighters (\#5) & Emotional range (anger, anxiety, depression, immoderation, self-consciousness, vulnerability) & Software Developer, Computer Engineer & Hacker (Pure) \\
  \hline
   Engineer (\#1) & Highest in openness in the facets of imagination and intellect. & Materials Engineer and Chemical Engineer. & Hipster \\
   \bottomrule
   \hline\label{tab:Table1}
\end{longtable}
\endgroup

For each founder and co-founder, we found the closest corresponding \emph{occupation tribe} for each based on personality similarity.  Then we tallied the founders within each cluster by tribe to reveal the level of coherence or the extent to which most founders within each group belonged to one \emph{occupation tribe}. 

This revealed three “purebred” clusters: \#0, \#2 and \#5, whose members are dominated by a single tribe (larger than 60\%). Thus these clusters represent and share personality attributes of these previously identified\cite{mccarthy2022occupation} occupation-personality tribes, which have the following known distinctive personality attributes:

\begin{itemize}
  \item \emph{Accomplishers (\#0)} — Organised \& outgoing. confident, down-to-earth, content, accommodating, mild-tempered \& self-assured.
  \item \emph{Leaders (\#2)} — Adventurous, persistent, dispassionate, assertive, self-controlled, calm under pressure, philosophical, excitement-seeking \& confident. 
  \item \emph{Fighters (\#5)} — Spontaneous and impulsive, tough, sceptical, and uncompromising. 

\end{itemize}

These labels also accord with the distribution of roles founders in each of these clusters hold. Accomplishers are often CEOs, CFOs or COOs while Fighters tend to be CTOs, CPOs and CCO. \emph{(See Extended Data Fig.~\ref{fig:ExtendedFig6} for more details)}.

We labelled these clusters with these tribe names, acknowledging that labels are somewhat arbitrary, based on our best interpretation of the data \emph{(See Extended Data Fig.~\ref{fig:ExtendedFig5} for more details)}. 

For the remaining three clusters \#1, \#3 and \#4, we can see they are “hybrids”, meaning that the founders within them come from a mix of different tribes, with no one tribe representing more than 50\% of the members of that cluster. However, the tribes with the largest share were noted as \#1 Experts; \#3 Fighters and \#4 Accomplishers. 

To label these three hybrid clusters, we examined the closest occupations to the median personality features of each cluster. We selected a name that reflected the common themes of these occupations, namely:

\begin{itemize}
  \item \emph{Engineers (\#1)} as the closest roles included Materials Engineers and Chemical Engineers. This is consistent with this cluster's personality footprint, which is highest in openness in the facets of imagination and intellect.
  \item \emph{Developers (\#3)} as the closest roles include Application Developers and related technology roles such as Business Systems Analysts and Product Managers. 
  \item \emph{Operators (\#4)} as the closest roles include service, maintenance and operations functions, including Bicycle Mechanic, Mechanic and Service Manager. This is also consistent with one of the key personality traits of high conscientiousness in the facet of orderliness and high agreeableness in the facet of humility for founders in this cluster. 

\end{itemize}

\begin{figure}
  {\includegraphics[width=0.4\textwidth, keepaspectratio]{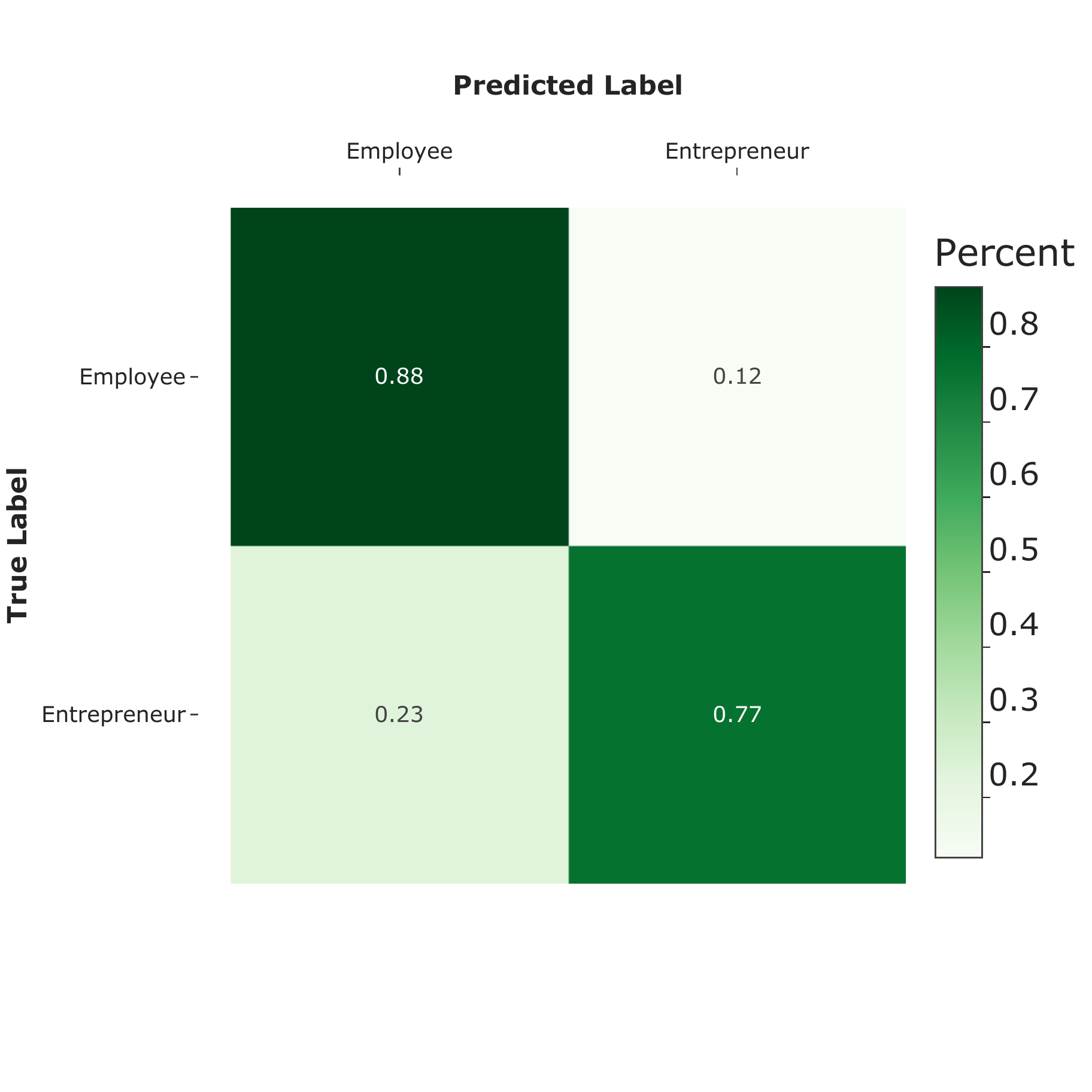}\label{fig:Figure2a}}
  {\includegraphics[width=0.4\textwidth]{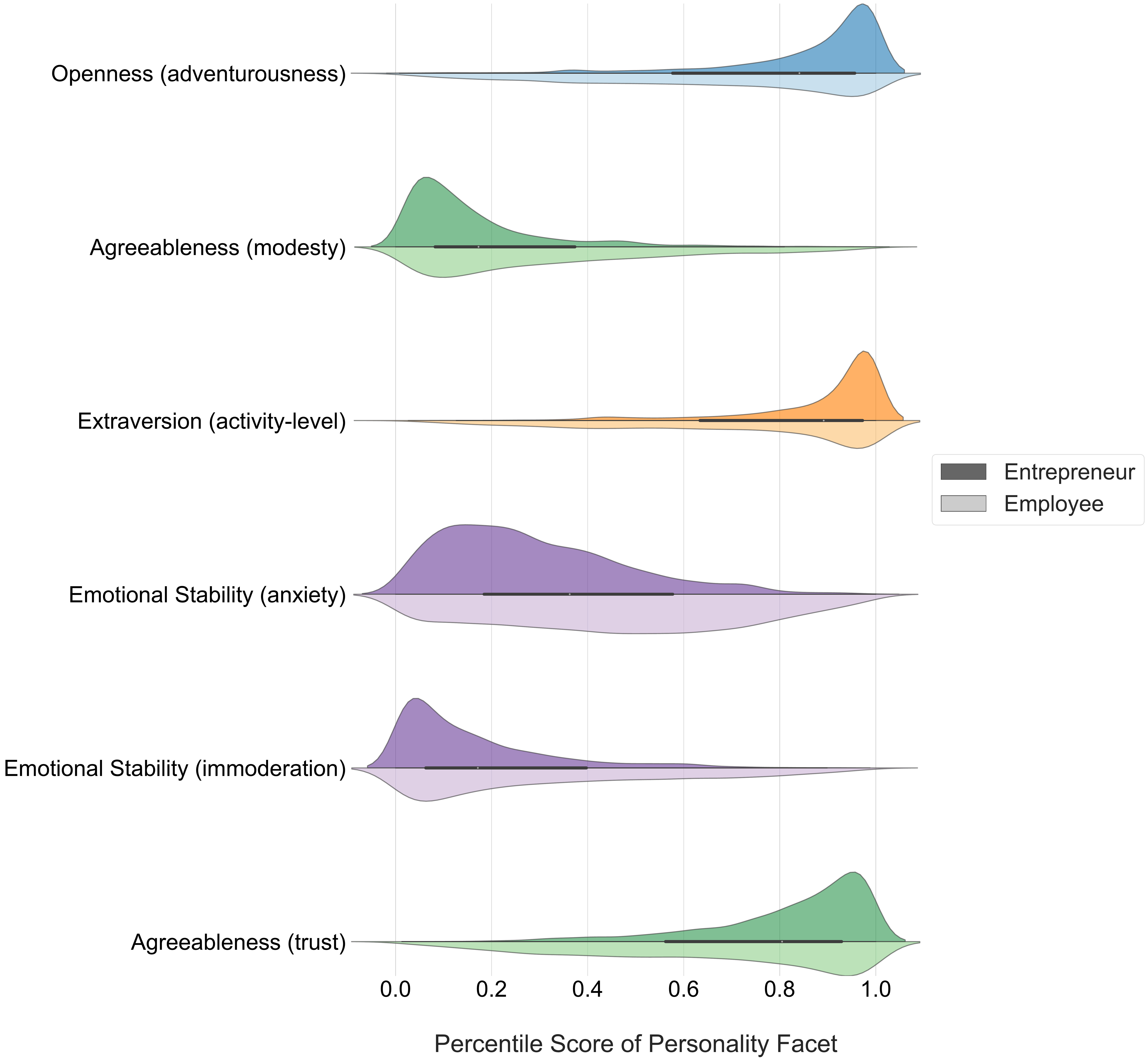}\label{fig:Figure2b}}\\[3ex]
  {\includegraphics[width=0.4\textwidth]{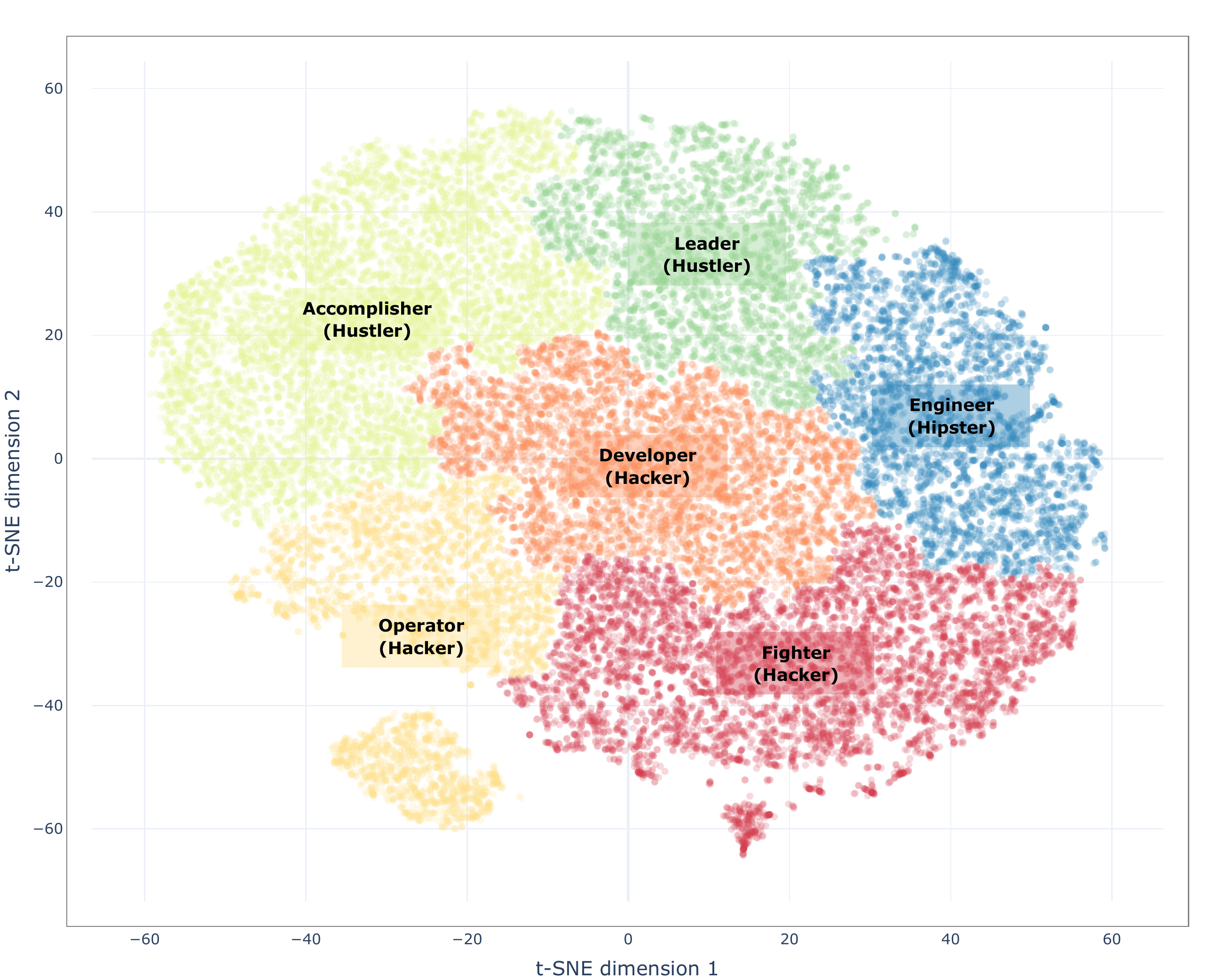}\label{fig:Figure2c}}
  {\includegraphics[width=0.4\textwidth]{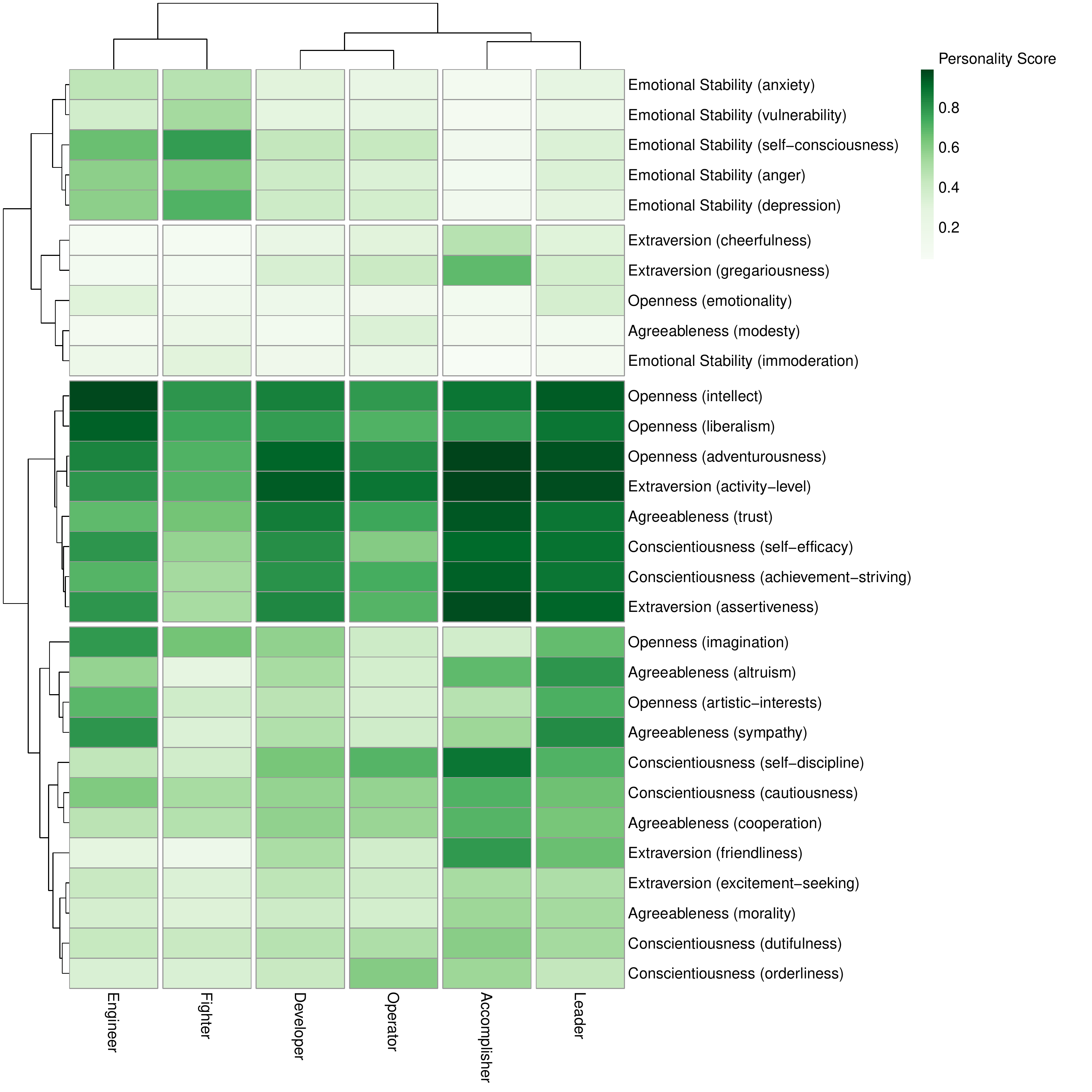}\label{fig:Figure2d}}
  \caption{\textbf{$|$ Founder-Level Factors of Startup Success.} \textbf{a}, Successful entrepreneurs differ from successful employees. They can be accurately distinguished using a classifier with personality information alone  \textbf{b}, Successful entrepreneurs have different Big 5 facet distributions, especially on adventurousness, modesty and activity level. \textbf{c}, Founders come in six different types: Fighters, Operators, Accomplishers, Leaders, Engineers and Developers (FOALED)  \textbf{d}, Each founder Personality-Type has its distinct facet footprint.}\label{fig:Figure2}
\end{figure}

Together, these six different types of startup founders (Fig.~\ref{fig:Figure2c}) represent a framework we call the FOALED model of founder types — an acronym of Fighters, Operators, Accomplishers, Leaders, Engineers and Developers. 

Each founder Personality-Type has its distinct facet footprint. Also, we observe a central core of correlated features that are high for all types of entrepreneurs, including intellect, adventurousness and activity level (Fig.~\ref{fig:Figure2d}).

\subsection*{Evidence for the “Hipster, Hacker, and Hustler” thesis}
By analysis of the six types of startup founders in our FOALED model within the broader Occupation-Personality landscape, we identify three types to be characterised as types of \emph{Hackers} (Fighters, Operators and Developers) and two as \emph{Hustlers} (Accomplishers and Leaders). The remaining type is different in personality to both Hackers and Hustlers. It is more of a subject matter expert whose insider field knowledge and problem-solving design strengths can be seen as a type of \emph{Hipster} (Engineer).  

\begin{figure}
  \includegraphics[width=0.4\textwidth]{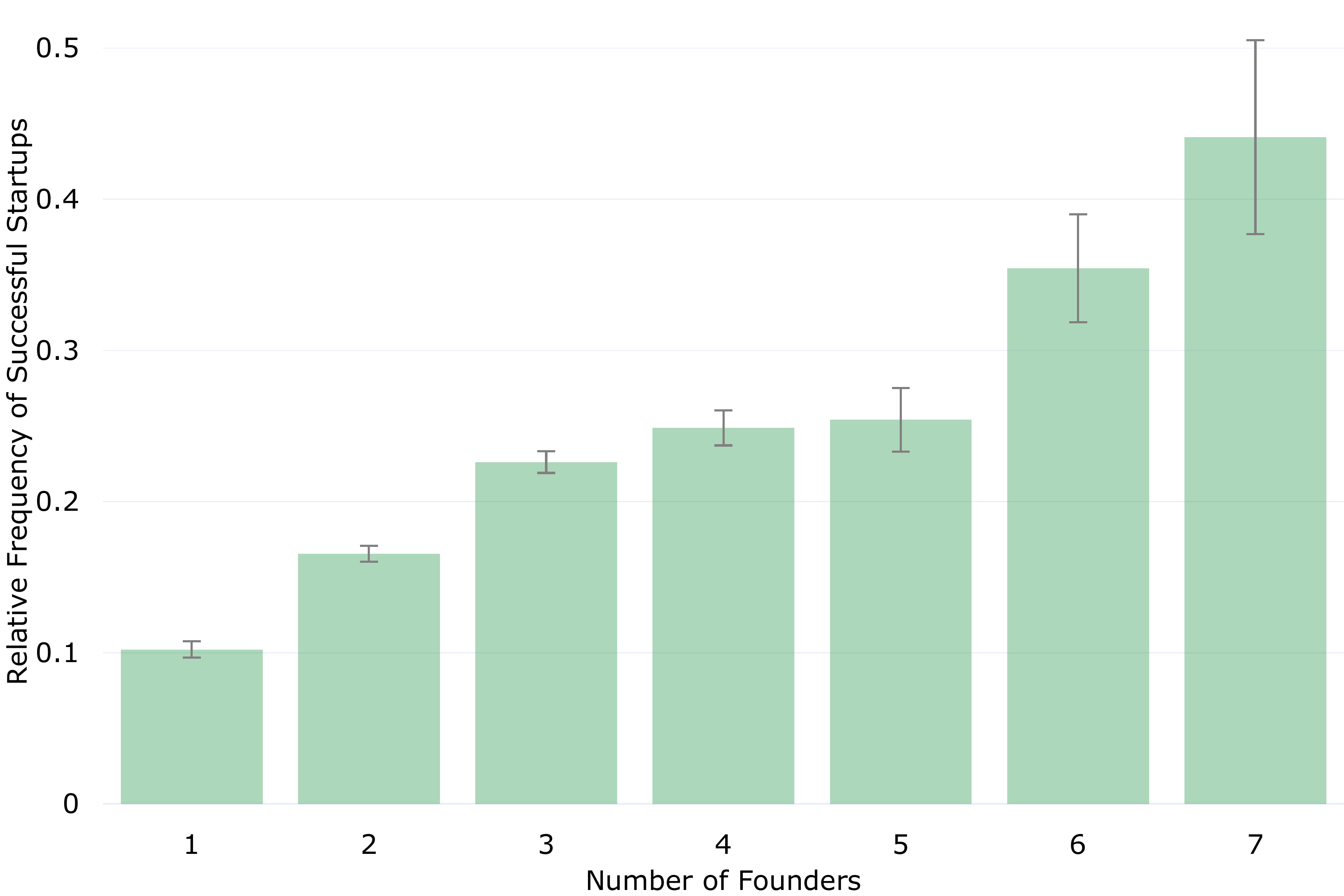}\label{fig:Figure3a}
  \includegraphics[width=0.6\textwidth]{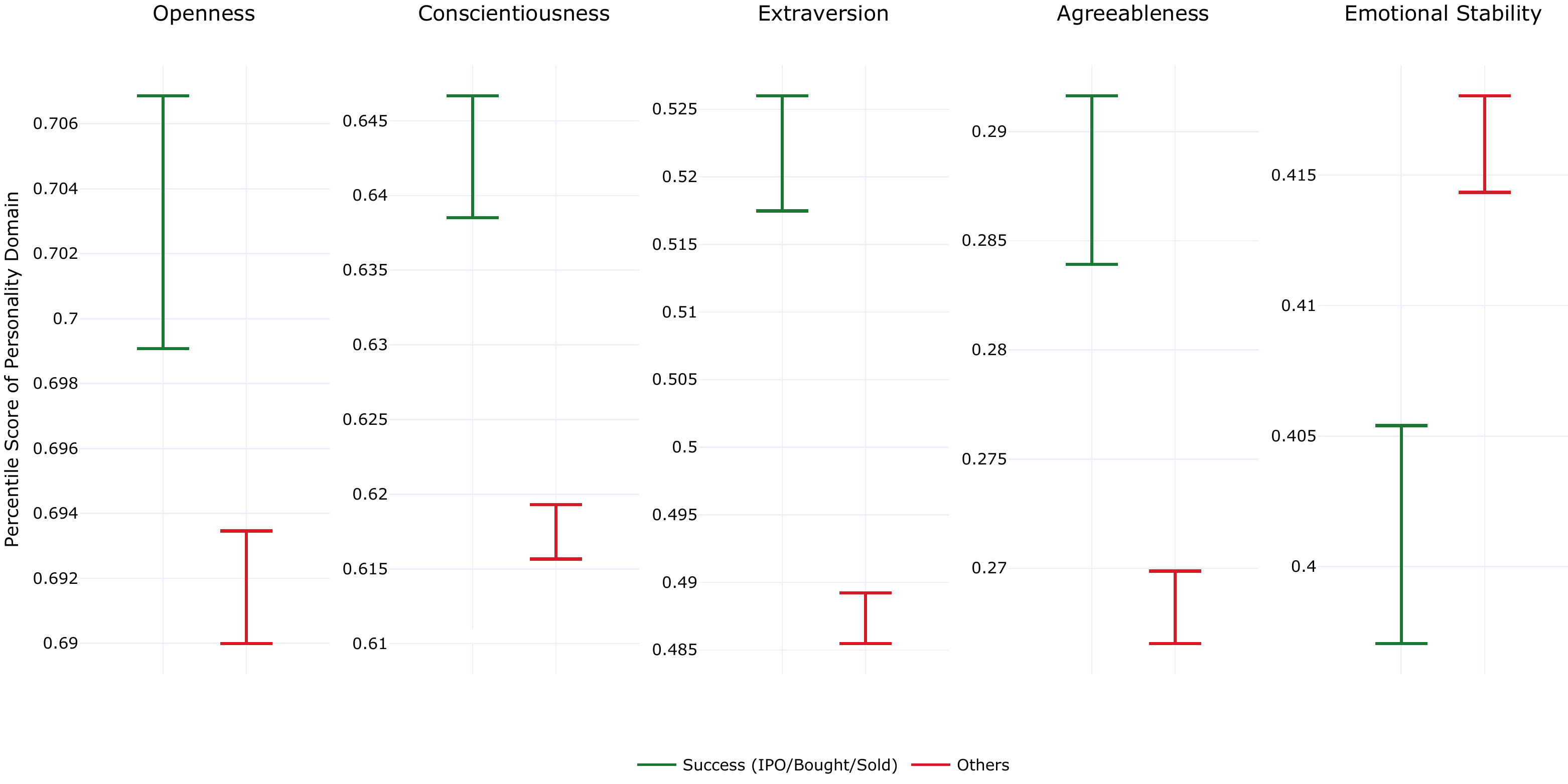}\label{fig:Figure3b}\\[3ex]
  \includegraphics[width=0.6\textwidth]{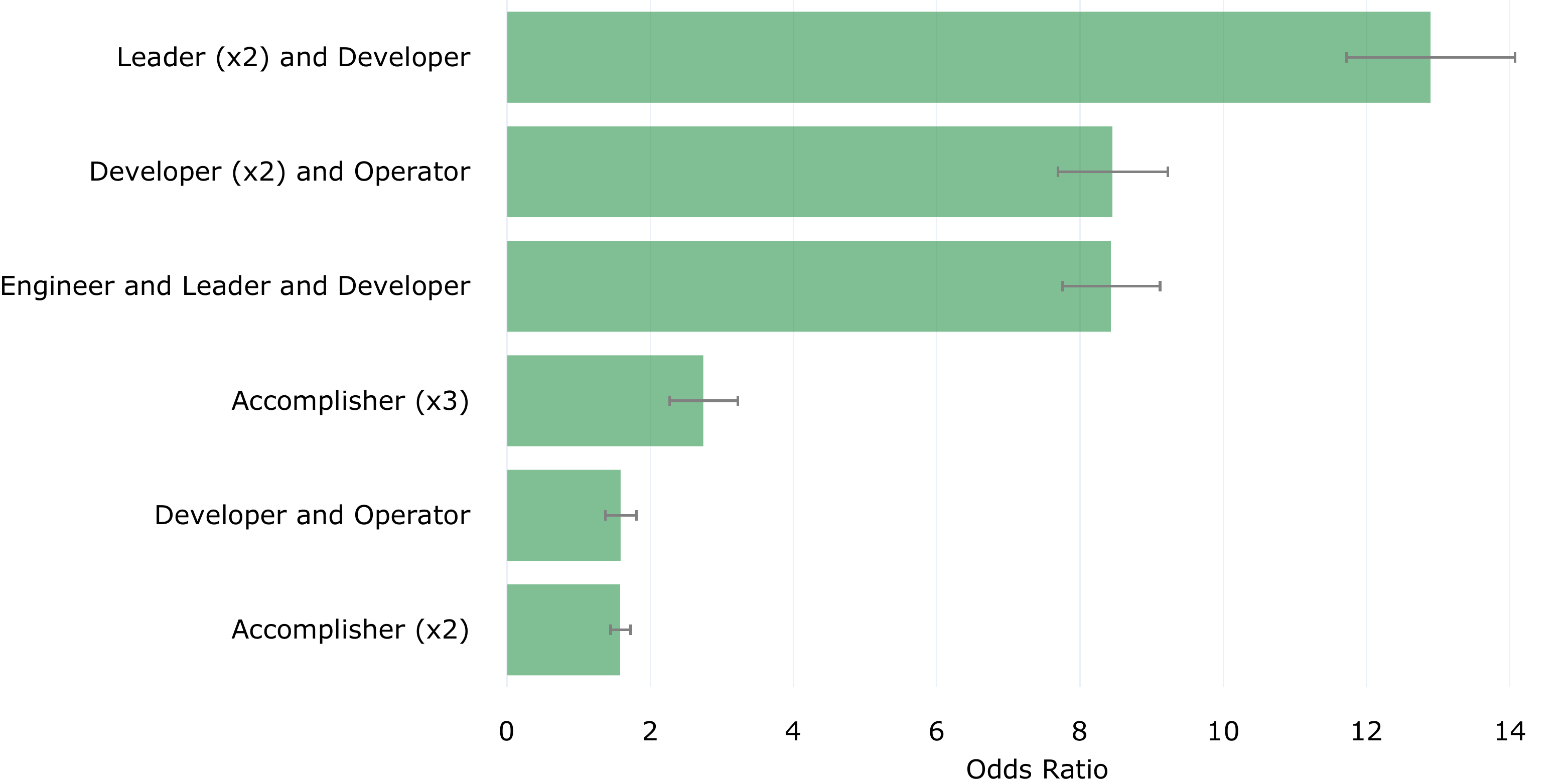}\label{fig:Figure3c}
  \caption{\textbf{$|$ The Ensemble Theory of Team-Level Factors of Startup Success.} \textbf{a}, Having a larger founder team elevates the chances of success. This can be due to multiple reasons, e.g., a more extensive network or knowledge base but also personality diversity. \textbf{b}, We show that joint personality combinations of founders are significantly related to higher chances of success. This is because it takes more than one founder to cover all beneficial personality traits that "breed" success. \textbf{c}, In our multifactor model, we show that firms with diverse and specific combinations of types of founders, including (Hipster, Hustler, and Hacker) have significantly higher odds of success.}\label{fig:Figure3}
\end{figure}

When we subsequently explored the combinations of personality types among founders and their relationship to the probability of the firm's success, adjusted for a range of other factors in a multi-factorial analysis, we found significantly increased chances of startup success for \emph{Hipster}, \emph{Hacker} and \emph{Hustler} foundation teams (Fig.~\ref{fig:Figure3c}).

\section*{Ensemble Theory of Success}

\subsection*{Definition of success}
The success of startups is uncertain, dependent on many factors and can be measured in various ways. Due to the likelihood of failure in startups, some large-scale studies have looked at which features predict startup survival rates\cite{antretter2018predicting} and others focus on fundraising from external investors at various stages\cite{dworak2022analysis}. Success for startups can be measured in multiple ways, such as the amount of external investment attracted, the number of new products shipped or the annual growth in revenue. But sometimes external investments are misguided, revenue growth can be short-lived, and new products may fail to find traction. 

The definition used by Bonaventura et al. \cite{bonaventura2020predicting}, namely that a startup either is acquired, acquires another company or has an initial public offering (IPO), sees any of these major capital liquidation events as a clear threshold signal that the company has matured from an early-stage venture to becoming or is on its way to becoming a mature company with clear and often significant business growth prospects.

Rather than looking at associations of any one factor of success, we use a quantitative multifactor analysis of success that incorporates a range of \emph{firm-level} factors such as where a startup is located, when it was founded and what industry it is in, combined with \emph{founder-level} factors such as the inferred Big 5 personality features in 30 dimensions for each founder and lastly \emph{founder-team level} factors that look at the number of founders and the permutations and combination of their personalities.  We look at these factors independently and in combination to explore their relative impacts on the likelihood of startup firm success. 

\subsection*{Factors associated with startup success }
Using multifactor analysis and a binary classification prediction model of startup success, we looked at many variables together and their relative influence on the probability of the success of startups. We looked at seven categories of factors through three lenses of firm-Level factors: 1) Location, 2) Industry, 3) Age of Startup; Founder-level factors: 4) Number of Founders, 5) Gender of Founders, 6) Personality characteristics of Founders and; lastly Team-level factors: 7) Founder-team personality combinations. 

The model performance and relative impacts on the probability of startup success of each of these categories of founders are illustrated in more detail in Extended Data Fig.~\ref{fig:ExtendedFig19} and in Extended Data Fig.~\ref{fig:ExtendedFig20} respectively.

In total, we considered over three hundred variables (n=323) and their relative significant association with success. 

\subsection*{Firm-level factors and success}
The first lens we looked through was at the \emph{firm-level}. Much of the previous literature on startups has been focused on firm-level or external factors and their influence on success\cite{zbikowski2021machine}. Startup success has been shown to relate to how much capital the startup has raised, how old it is and what industry it is in, among other things\cite{corea2021hacking}. 

Here we show startup success is influenced strongly by its location (firms from Japan, Scandinavia, USA, France, and Germany are more likely to be successful than those from Turkey, Argentina, Mexico or other countries); industry (firms in Payment Systems and Privacy \& Security are most successful) and a company's age (more details in the SI).

\subsection*{Founder-level factors and success}
The second lens we looked through was that of \emph{founder-level} factors or those internal to the firm, i.\,e. the personality features of founders and their association with success. Our modelling shows firms with multiple founders are more likely to succeed, as illustrated in Fig.~\ref{fig:Figure3a}), which shows firms with three or more founders are more than twice as likely to succeed as solo-founded startups. This finding is consistent with investors' advice to founders and previous studies\cite{klotz2014new}. (We also noted that some types of additional founders increase the probability of success more than others as shown in Extended Data Fig.~\ref{fig:ExtendedFig14} and Extended Data Fig.~\ref{fig:ExtendedFig15}). 

Access to more extensive networks and capital could explain the benefits of having more founders. Still, as we find here, it also offers a greater diversity of combined personalities, which naturally provides a broader range of maximum traits. So, for example, one founder may be more open and adventurous, and another could be highly agreeable and trustworthy, thus potentially complementing each other's particular strengths associated with startup success. 

The benefits of larger and more personality-diverse foundation teams can be seen in the apparent differences between successful and unsuccessful firms based on their combined Big 5 personality team footprints, illustrated in Figure~\ref{fig:Figure3b}). Here maximum values within each startup for each Big 5 trait for any of its cofounders are mapped, and the spread of these between successful firms — those who have IPOed, been acquired or acquired another firm and the other firms are shown.

\subsection*{Team-level factors and success}
Lastly, we considered \emph{team-level} factors — founder team personality combinations and how they related to startup success.   

We found that ten combinations of founders with different personality types were significantly correlated with greater chances of startup success when accounting for other variables in the model. The coefficient of each of these factors is illustrated concerning other features that were also found to be significantly associated with success in Figure~\ref{fig:Figure3c} (see Supplementary Figure~\ref{fig:ExtendedFig20} for more details on the performance of modelling).

Three combinations of trio-founder companies were more than twice as likely to succeed than other combinations, namely teams with:

\begin{itemize}
  \item A \emph{Leader} and two \emph{Developers} (a hustler and two hackers)
  \item An \emph{Operator} and two \emph{Developers} (three hackers of two different types) 
  \item An \emph{Engineer}, \emph{Leader} and \emph{Developer} (a hipster, hustler and hacker) 

\end{itemize}

The last of these aligns with and provides evidence for the \emph{Hipster, Hustler \& Hacker} hypothesis as well as a commonality of \emph{Developers} or “purebred hackers” in all three of the most successful combinations. 

\section*{Discussion}
Startups are one of the key mechanisms for brilliant ideas to become solutions to some of the world’s most challenging economic and social problems. Examples include the Google search algorithm, disability technology startup Fingerwork’s touchscreen technology that became the basis of the Apple iPhone, or the Biontech mRNA technology that powered Pfizer’s COVID-19 vaccine.

We have shown that founders' personalities and the combination of personalities in the founding team of a startup have a material and significant impact on its likelihood of success. We have also shown that successful startup founders' personality traits are significantly different from those of successful employees - so much so that a simple predictor can be trained to distinguish between employees and entrepreneurs with more than 80\% accuracy using personality trait data alone. 

Just as \emph{occupation-personality} maps derived from data based on people already successful in those roles can provide career-guidance tools, so too can data on successful entrepreneurs' personality traits help others decide whether to become a founder may be a good choice for them.

We have learnt through this research that there is not one type of ideal “entrepreneurial” personality but six different types. Many successful startups have multiple co-founders with a combination of these different personality types. 

Startups are, to a large extent, a team sport; as such, diversity and complementarity of personalities matter in the foundation team. It has an outsized impact on the company's likelihood of success. While all startups are high risk, the risk becomes lower with more founders, particularly if they have distinct personality traits. Our work demonstrates the benefits of diversity among the founding team of startups. Greater awareness of these benefits may help create more resilient startups capable of more significant innovation and impact. 

\subsection*{Biases and Limitations}
While each is large and comprehensive, there are some known and likely sample biases in the principal data sources used (namely Crunchbase, Twitter and LinkedIn).

Crunchbase is the principal public chronicle of Venture Capital funding, and so there is some likely sample bias toward:

\begin{itemize}
  \item Startup companies that are funded externally. Self-funded or bootstrapped companies are less likely to be represented in Crunchbase. 
  \item Technology companies, as that is Crunchbase’s roots.
  \item Multifounder companies. As it’s a public social record, companies with multiple founders are likely better represented in Crunchbase than those with one founder.
  \item Male founders. Like the technology industry itself, founders represented in Crunchbase are overwhelmingly male. Although the representation of female founders is now double that of the mid-2000s, women still represent less than 25\% of the sample. (See Extended Data Fig. ~\ref{fig:ExtendedFig18} for more detail of how this manifests in the data):
  \item Companies that succeed.  Companies that fail, especially those that fail early, are likely to be less represented in the data.

\end{itemize}

Samples were also limited to those whose founders are active on Twitter, which adds additional selection biases. For example, Twitter users typically are younger, more educated and have a higher median income\cite{duggan2015demographics}.

In addition to sampling biases within the data, there are also significant historical biases in startup culture. For many aspects of the entrepreneurship ecosystem, women, for example, are at a disadvantage\cite{brush2019gendered}. Male-founded companies have historically dominated most startup ecosystems worldwide, representing the majority of founders and the overwhelming majority of venture capital investors.  As a result, startups with women have historically attracted significantly fewer funds\cite{kanze2018we}, in part due to the male bias among venture investors, although this is now changing, albeit slowly\cite{fan2022startup}.

\subsection*{Opportunities and Future research questions}
The global startup ecosystem is evolving, bringing a variety of questions for the dynamics of startups. For instance:

\begin{itemize}
  \item Will the recent growing focus on promoting and investing in female founders change the nature, composition and dynamics of startups \textit{and} their personalities? 
  \item Will the growth of startups outside of the United States change what success looks like to investors and hence the role of different personality traits and their association to diverse success metrics?
  \item Many of today’s most renowned entrepreneurs are either \emph{Baby Boomers} (Gates, Branson, Bloomberg) or \emph{Generation Xers} (Benioff, Cannon-Brooks, Musk). However, as we can see, personality is both a predictor and driver of success in entrepreneurship. Will generation-wide differences in personality and outlook affect startups and their success?

\end{itemize}

The findings of this research have natural extensions and applications beyond startups, such as for new projects within large established companies. While not technically startups, many large enterprises and industries such as construction, engineering and the film industry rely on forming new project-based, cross-functional teams that are often new ventures and share many characteristics of startups.

There is also potential for extending this research in other settings in government, NGOs and within research itself. In scientific research, for example, team diversity in terms of age, ethnicity and gender has been shown to be predictive of impact, and personality diversity may be another critical dimension\cite{alshebli2018preeminence}.

This study demonstrates that successful startup founders have significantly different personalities than many successful employees. It also shows that many factors influence startup success. The methods and data described here reveal that \emph{firm-level} factors such as the startup's context within geography (where it is located), the economy (which industry it addresses), and timing (when it was founded and how old it is) all have a significant influence of the likelihood of firm success. In addition to these more well-understood factors, we showed that a range of \emph{founder-level} factors, notably the character traits of its founders, as revealed by their personality features, have a significant impact on a startup's likelihood of success.  Lastly, we looked at \emph{team-level} factors and discovered in a multifactor analysis that personality-diverse teams have the most considerable impact of all those examined on the probability of a startup's success.

\printbibliography


\section*{Methods}

\subsection*{Data Sources}
\textbf{Entrepreneurs Only (EO) Dataset.} Data about the founders of startups were collected from Crunchbase (Table~\ref{tab:Table2}), an open reference platform for business information about private and public companies, primarily early-stage startups.  It is one of the largest and most comprehensive data sets of its kind and has been used in over 100 peer-reviewed research articles about economic and managerial research.

Crunchbase contains data on over two million companies - mainly startup companies and the companies who partner with them, acquire them and invest in them, as well as profiles on well over one million individuals active in the entrepreneurial ecosystem worldwide from over 200 countries and spans. While Crunchbase started in the technology startup space, it now covers all sectors, specifically focusing on entrepreneurship, investment and high-growth companies. 

While Crunchbase contains data on over one million individuals in the entrepreneurial ecosystem, some are not entrepreneurs or startup founders but play other roles, such as investors, lawyers or executives at companies that acquire startups.  To create a subset of only entrepreneurs, we selected a subset of 32,732 who self-identify as founders and co-founders (by job title) and who are also publicly active on the social media platform Twitter. We also removed those who also are venture capitalists to distinguish between investors and founders. 

We selected founders active on Twitter to be able to use natural language processing to infer their Big 5 personality features using an open-vocabulary approach shown to be accurate in the previous research by analysing users' unstructured text, such as Twitter posts in our case.  For this project, as with previous research (Kern et al. 2019), we employed a commercial service, IBM Watson Personality Insight, to infer personality facets. This service provides raw scores and percentile scores of Big Five Domains (Openness, Conscientiousness, Extraversion, Agreeableness and Emotional Stability) and the corresponding 30 Subdomain or facets. In addition, the public content of Twitter posts was collected, and there are 32,732 profiles that each had enough Twitter posts (more than 150 words) to get relatively accurate personality scores (less than 12.7\% Average Mean Absolute Error). 

The “Entrepreneurs Only” (EO) dataset is analysed in combination with other data about the companies they founded to explore questions around the nature and patterns of personality traits of entrepreneurs and the relationships between these patterns and company success.

For the multifactor analysis, we cleaned EO the data filtering by a number of factors to ensure the sample was robust and consistent. More details on this data wrangling is included in Extended Data Fig.~\ref{fig:ExtendedFig7} and Extended Data Fig.~\ref{fig:ExtendedFig8}. 

\begingroup
\footnotesize
\singlespacing
\begin{longtable}{ C{3cm}  C{2.5cm}  C{2.5cm}  C{2.5cm}  C{3cm}}
\caption[$|$ Summary of the basic information of the Entrepreneurs Only (EO) dataset]{
    \textbf{$|$ Summary of the basic information of the Entrepreneurs Only (EO) dataset}
    the number of founders and associated startups in population, how many countries those startups are across, and the time span the data collected covers, the number of features included. ).
}\\
    \hline
    \toprule
    \small
    \makecell{\textbf{Founders with} \\ \textbf{Personality Data}} & \makecell{\textbf{Associated} \\  \textbf{Startups}}  & \makecell{\textbf{Countries}} & \makecell{\textbf{Date Range}} & \makecell{\textbf{Founders Individual} \\ \textbf{Features}}\\
    \hline
    \endhead
   32,732 & 23,292 & 215 & 2008-2021 & 100 \\
   \bottomrule
   \hline\label{tab:Table2}
\end{longtable}
\endgroup

\textbf{Successful Entrepreneurs and Successful Employees (SESE) Dataset.} The EO data set contains two categories of Founders: those that have raised funds or attracted external investment to their companies or Funded Founders (n=17,057) and those who have not - Unfunded Founders (n=16,675). The attraction of a significant investment from outside, especially from specialist venture capitalists, is seen as one measure that indicates a startup has had some degree of success or, at the very least, shows promise of future success. Therefore, we filtered the EO Funded Founders by those whose companies had attracted more than US\$100k in investment to create a reference set of Successful Entrepreneurs (n=4,400).

Most company founders also adopt regular occupation titles such as CEO or CTO. Many founders will be Founder and CEO or Co-founder and CTO. While founders are often CEOs or CTOs, the reverse is not necessarily true, as many CEOs are professional executives that were not involved in the establishment or ownership of the firm.

To create a control group of \emph{Successful Employees}, who are \emph{not also entrepreneurs} or very unlikely to be of have been entrepreneurs, we leveraged the fact that while some occupational titles like CEO, CTO and \emph{Public Speaker} are commonly shared by founders and co-founders, some others such as \emph{Cashier}, \emph{Zoologist} and \emph{Detective} very rarely co-occur with founder or co-founder. Using data from LinkedIn, we created an \emph{Entrepreneurial Occupation Index (EOI)} based on the ratio of entrepreneurs for each of the 624 occupations used in a previous study of occupation-personality fit. It was calculated based on the percentage of all people working in the occupation from LinkedIn compared to those who shared the title Founder or Co-founder (See SI for more detail). A reference set of \emph{Successful Employees} (n=6,685) was then selected across 112 different occupations with the lowest propensity for entrepreneurship (less than 0.5\% EOI) from a large corpus of Twitter users with known occupations, also from the previous occupational-personality fit study (PX McCarthy and others, 2022). 

The \emph{Successful Entrepreneurs} and \emph{Successful Employees} were combined to create the SEE dataset, which was used to test whether it may be possible to distinguish successful entrepreneurs from successful employees based on the different patterns of personality traits alone.

\subsection*{Hierarchical Clustering}
We applied a number of clustering techniques and tests to the personality vectors of the EO data set to determine if there are natural clusters and, if so, how many are the optimum number. 

Firstly, to determine if there is a natural typology to founder personalities, we applied the \textbf{Hopkins statistic} - a statistical test we used to answer whether the “EO” dataset contains inherent clusters. It measures the clustering tendency based on the ratio of the sum of distances of real points within a sample of the “EO” dataset to their nearest neighbours and the sum of distances of randomly selected artificial points from a simulated uniform distribution to their nearest neighbours in the real “EO” dataset. The ratio measures the difference between the “EO” data distribution and the simulated uniform distribution, which tests the randomness of the data. The range of Hopkins statistics is from 0 to 1. Where the scores are close to 0, 0.5 and 1, respectively, this indicates whether the dataset is uniformly distributed, randomly distributed or highly clustered. 

To cluster the founders by personality facets, we used \textbf{Agglomerative Hierarchical Clustering (AHC)} - a bottom-up approach that treats an individual data point as a singleton cluster and then iteratively merges pairs of clusters until all data points are included in the single big collection. Ward’s linkage method is used to choose the pair of clusters for minimising the increase in the within-cluster variance after combining. AHC was widely applied to clustering analysis since a tree hierarchy output is more informative and interpretable than K-means. Dendrograms were used to visualise the hierarchy to provide the perspective of the optimal number of clusters. The heights of the dendrogram represent the distance between groups, where the lower heights represent more similar groups of observations. A horizontal line through the dendrogram was drawn to distinguish the number of significantly different clusters with higher heights. However, as it is not possible to determine the optimum number of clusters from the dendrogram, we applied other clustering performance metrics to analyse the optimal number of clusters.

A range of \textbf{Clustering performance metrics} were used to help determine the optimal number of clusters in the dataset after an obvious clustering tendency was confirmed. The following metrics were implemented to comprehensively evaluate the differences between within-cluster and between-cluster distances: Dunn Index, Calinski-Harabasz Index, Davies-Bouldin Index and Silhouette Index. The Dunn Index measures the ratio of the minimum inter-cluster separation and the maximum intra-cluster diameter. At the same time, the Calinski-Harabasz Index improves the measurement of the Dunn Index by calculating the ratio of the average sum of squared dispersion of inter-cluster and intra-cluster. The Davies-Bouldin Index simplifies the process by treating each cluster individually, which compares the sum of the average distance among intra-cluster data points to its cluster centre of two separate clusters with the distance between their centre points. Finally, the Silhouette Index is the overall average of the silhouette coefficients for each sample. The coefficient measures the similarity of the data point to its cluster compared with the other clusters. Higher scores of the Dunn, Calinski-Harabasz and Silhouette Index and a lower score of the Davies-Bouldin Index indicate better clustering configuration.

\subsection*{Classification Modelling}
\textbf{Classification algorithms.} To obtain a comprehensive and robust conclusion, we explored the following classifiers: Naïve Bayes, Elastic Net regularisation, Support Vector Machine, Random Forest, Gradient Boosting and Stacked Ensemble. The Naïve Bayes classifier is a probabilistic algorithm based on Bayes’ theorem with assumptions of independent features and equiprobable classes. Compared with other more complex classifiers, it saves computing time for large datasets and performs better if the assumptions hold. However, in the real world, those assumptions are generally violated. Elastic Net regularisation combines the penalties of Lasso and Ridge to regularise the Logistic classifier. It eliminates the limitation of multicollinearity in the Lasso method and improves the limitation of feature selection in the Ridge method. Even though Elastic Net is as simple as the Naïve Bayes classifier, it is more time-consuming. The Support Vector Machine (SVM) aims to find the ideal line or hyperplane to separate successful entrepreneurs and employees in this study. The dividing line can be non-linear based on a non-linear kernel, such as the Radial Basis Function Kernel. Therefore, it performs well on high-dimensional data while the “right” kernel selection needs to be tuned. Random Forest (RF) and Gradient Boosting Trees (GBT) are ensembles of decision trees. All trees are trained independently and simultaneously in RF, while a single new tree is trained each time and is corrected by previously trained trees in GBT. RF is a more robust and simple model since it does not have many hyperparameters to tune. GBT optimises the objective function and learns a more accurate model since there is a successive learning and correction process. Stacked Ensemble combines all existing classifiers through a Logistic Regression. Better than bagging with only variance reduction and boosting with only bias reduction, the ensemble leverages the benefit of model diversity with both lower variance and bias. All the above classification algorithms distinguish successful entrepreneurs and employees based on the personality matrix.  

\textbf{Evaluation metrics.} A range of evaluation metrics comprehensively explains the performance of a classification prediction. The most straightforward metric is accuracy, which measures the overall portion of correct predictions. It will mislead the performance of an imbalanced dataset. The F1 score is better than accuracy by combining precision and recall and considering the False Negatives and False Positives. Specificity measures the proportion of detecting the true negative rate that correctly identifies employees, while Positive Predictive Value (PPV) calculates the probability of accurately predicting successful entrepreneurs. Area Under the Receiver Operating Characteristic Curve (AUROC) determines the capability of the algorithm to distinguish between successful entrepreneurs and employees. A higher value means the classifier performs better on separating classes.

\textbf{Feature importance.} To further understand and interpret the classifier, it is critical to identify variables with significant predictive power on the target. Feature importance of tree-based models measures Gini importance scores for all predictors, which evaluate the overall impact of the model after cutting off the specific feature. The measurements consider all interactions among features. However, it does not provide insights into the directions of impacts since the importance only indicates the ability to distinguish different classes.

\textbf{Statistical analysis.} T-test, Cohen’s D and two-sample Kolmogorov-Smirnov test are introduced to explore how the mean values and distributions of personality facets between successful entrepreneurs and employees differ. The T-test is applied to determine whether the mean of personality facets of two group samples are significantly different from one another or not. The facets with significant differences detected by the hypothesis testing are critical to separate the two groups. Cohen’s d is to measure the effect size of the results of the previous t-test, which is the ratio of the mean difference to the pooled standard deviation. A larger Cohen’s d score indicates that the mean difference is greater than the variability of the whole sample. Moreover, it is interesting to check whether the probability distributions of personality facets of the two groups are from the same distribution through the two-sample Kolmogorov-Smirnov test. There is no assumption about the distributions, but the test is sensitive to deviations near the centre rather than the tail. 

\section*{Privacy and ethics}
The focus of this research is to provide high-level insights about groups of startups, founders and types of founder teams rather than on specific individuals or companies. While we used unit record data from the publicly available data of company profiles from \emph{Crunchbase}, we removed all identifiers from the underlying data on individual companies and founders and generated aggregate results, which formed the basis for our analysis and conclusions.

\section*{Data and Code Availability}
A dataset which includes only aggregated statistics about the success of startups and the factors that influence is released as part of this research. Underlying data for all figures and the code to reproduce them are also available. 

\section*{Acknowledgements}
We thank Gary Brewer from \emph{BuiltWith}; Leni Mayo from \emph{Influx}, Rachel Slattery from \emph{TeamSlatts} and Daniel Petre from \emph{AirTree Ventures} for their ongoing generosity and insights about startups, founders and venture investments. We also thank Tim Li from \emph{Crunchbase} for advice and liaison regarding data on startups and Richard Slatter for advice and referrals in \emph{Twitter}. 

\section*{Author contributions}
All authors designed research; All authors analysed data and undertook investigation; FB and FS led multi-factor analysis; PM, XG and MAR led the founder/employee prediction; MLK led personality insights; XG collected and tabulated the data; XG, FB, and FS created figures; XG created final art, and all authors wrote the paper.

\section*{Competing interests}
The authors have declared that no competing interests exist.



\newpage
\appendix
\renewcommand{\figurename}{Extended Data Fig.}
\renewcommand{\tablename}{Extended Data Table}
\setcounter{figure}{0}
\setcounter{table}{0}

\section*{Supplementary Information}

\begin{figure}[htbp]
	\centering
	\includegraphics[width=0.76\linewidth]{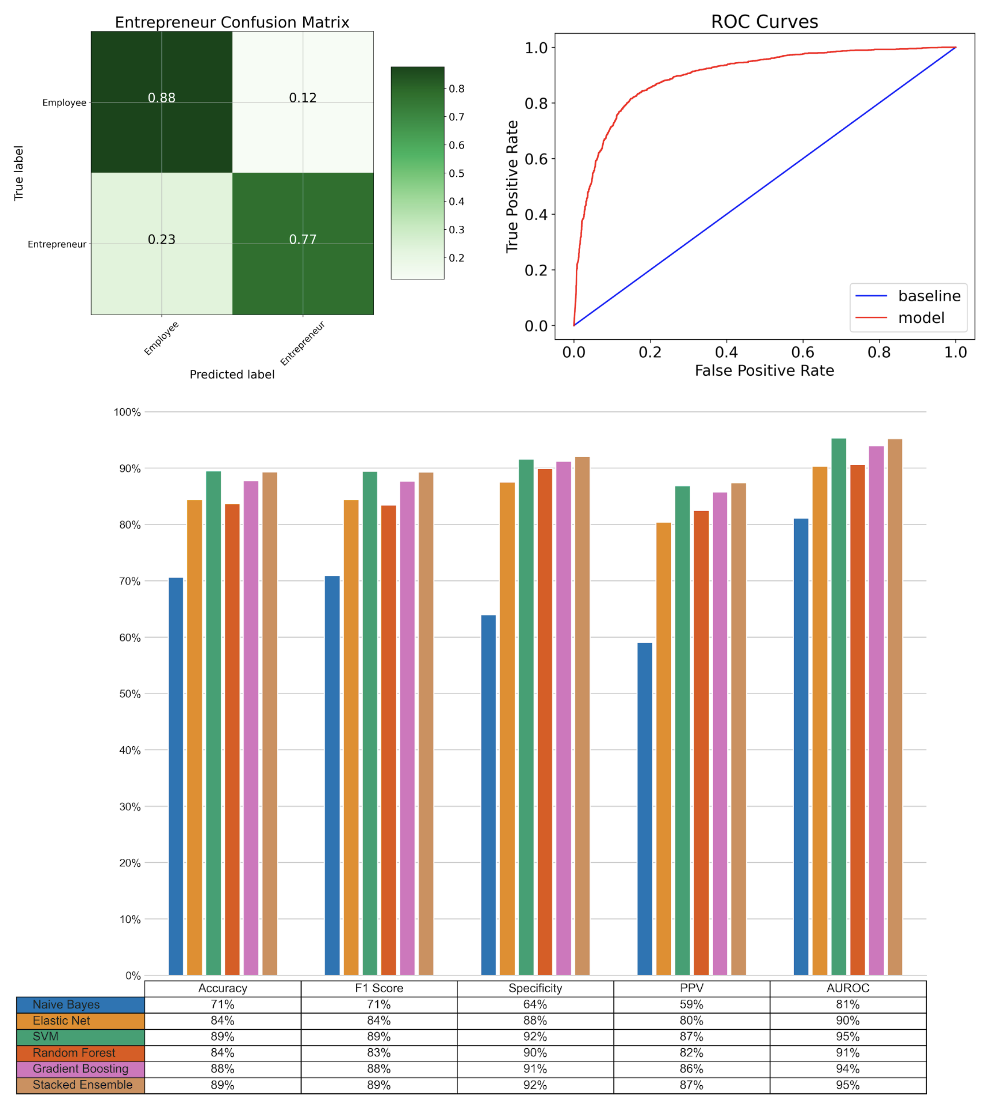}
	\vskip2em
	\caption{
		\textbf{$|$ Entrepreneurial Prediction Performance} 
		The ROC score of the train set is 0.94, while the score of the test set is 0.9. The confusion matrix also demonstrates high accuracy. On the test set, the model could 88\% correctly predict the entrepreneurs and 77\% correctly predict the employees.		
	}
	\label{fig:ExtendedFig1}
\end{figure}

\newpage
\begin{figure}[htbp]
	\centering
	\includegraphics[width=0.76\linewidth]{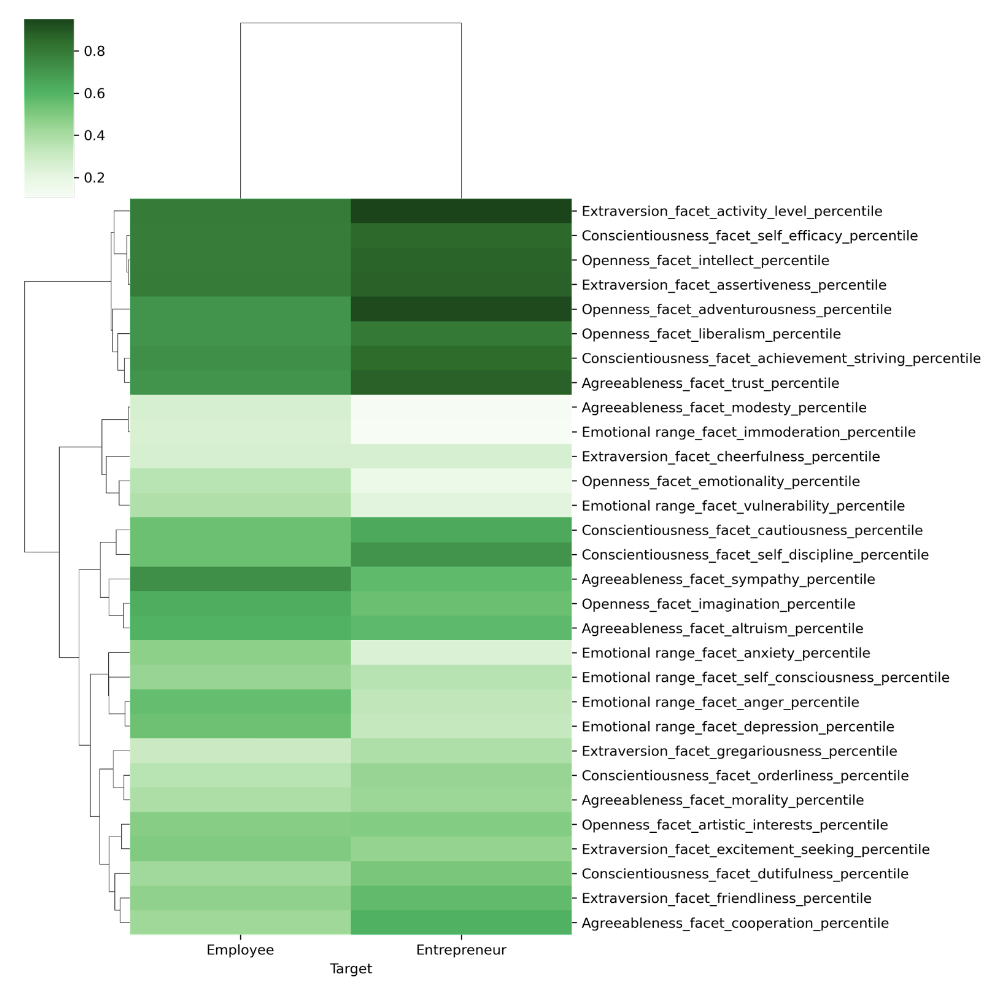}
	\vskip2em
	\caption{
		\textbf{$|$ Entrepreneurial Personality Facets} 
		The heatmap visualises the median values of 30 facets of two samples, which intuitively and comprehensively compare all personality traits and patterns. From the heatmap, it could be observed that the difference in median values of the two groups on adventurousness (Openness), activity level (Extraversion) and modesty (Agreeableness) are relatively large.		
	}
	\label{fig:ExtendedFig2}
\end{figure}

\newpage
\begin{figure}[htbp]
	\centering
	\includegraphics[width=0.76\linewidth]{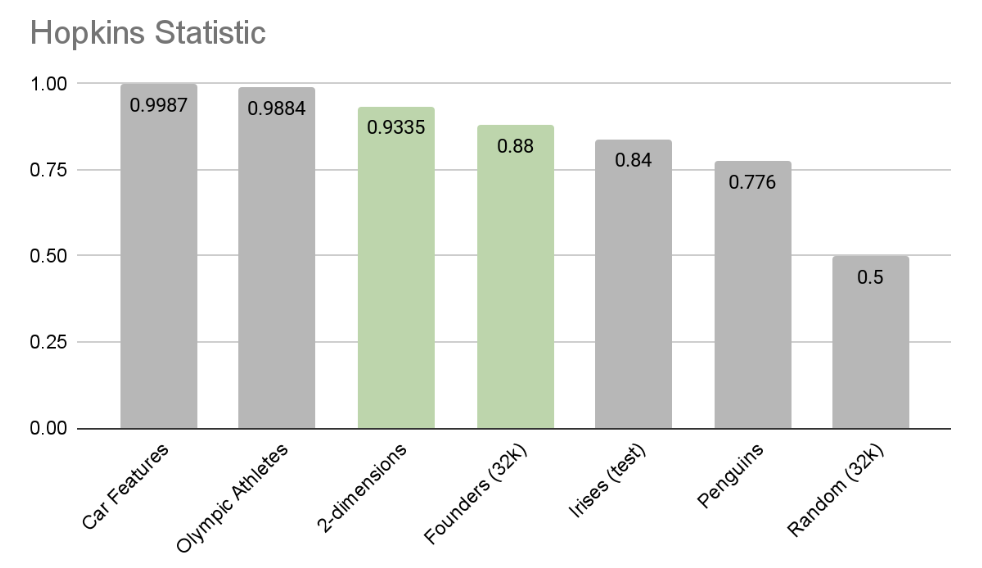}
	\vskip2em
	\caption{
		\textbf{$|$ Clustering Tendency Startup Founders by Personality Feaures} 
		Hopkins index (above) measures of the Founders inferred personality traits data, and 2-dimensions data (dimensionality reduction of Founders data) compares favourably to other famous test data sets (Irises; Penguins, Olympic Athletes) that are known to have explicit classes in the data — different breeds of Penguins, various species of Iris flowers and Olympic athletes who have qualified for different categories of events.		
	}
	\label{fig:ExtendedFig3}
\end{figure}

\newpage
\begin{figure}[htbp]
	\centering
	\includegraphics[width=0.76\linewidth]{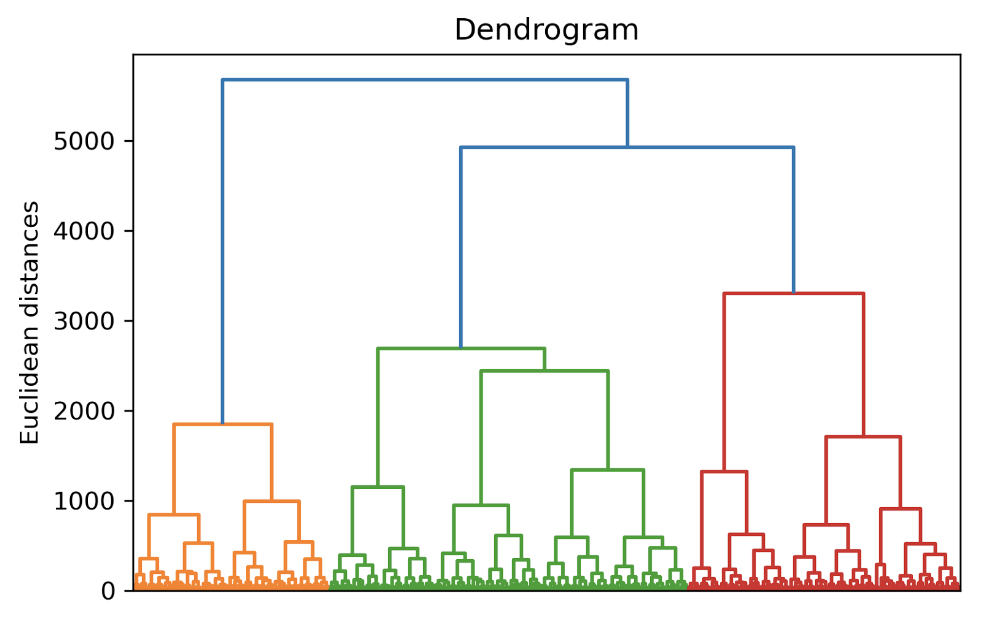}
	\vskip2em
	\caption{
		\textbf{$|$ Hierarchical Clustering of Startup Founders by Personality Feaures} 
		Hierarchical clustering was used to model potential clusters of startup founders by their personality features. 		
	}
	\label{fig:ExtendedFig4}
\end{figure}

\newpage
\begin{figure}[htbp]
	\centering
	\includegraphics[width=0.76\linewidth]{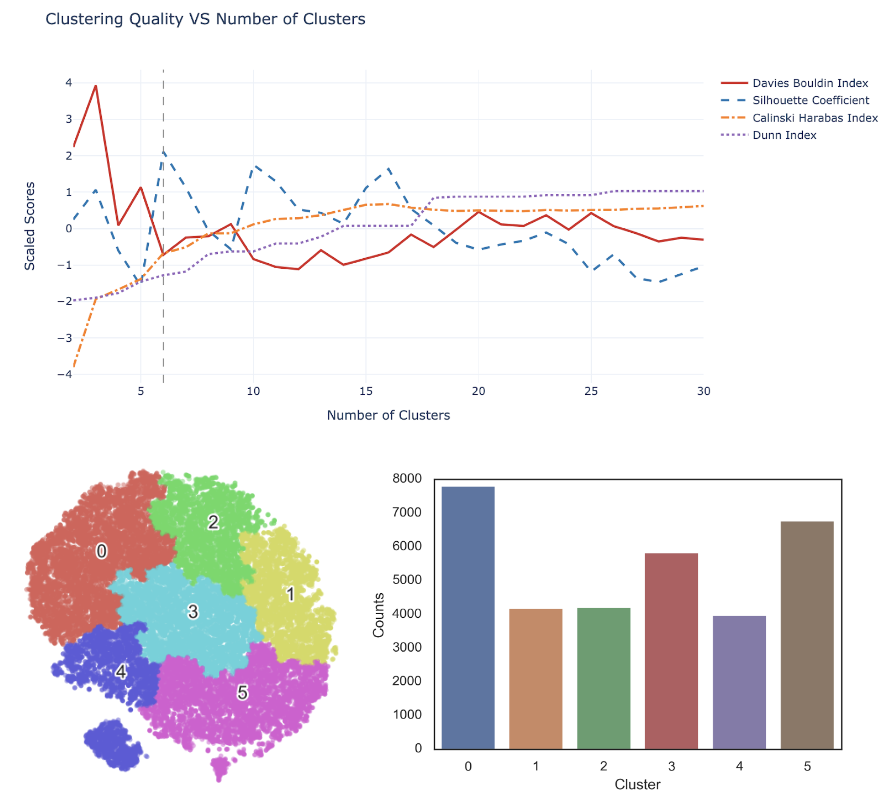}
	\vskip2em
	\caption{
		\textbf{$|$ Optimum Number of Clusters of Types of Founders} 
		Four different clustering quality indices were used to determine that there are optimally six clusters of startup founders. We labelled each of these \#0, \#1, \#2, \#3, \#4 and \#5 and produced the dendrogram, bar charts, and T-SNE plot above to demonstrate the hierarchy, adjacencies and distribution of all six clusters. 		
	}
	\label{fig:ExtendedFig5}
\end{figure}

\newpage
\begin{figure}[htbp]
	\centering
	\includegraphics[width=0.76\linewidth]{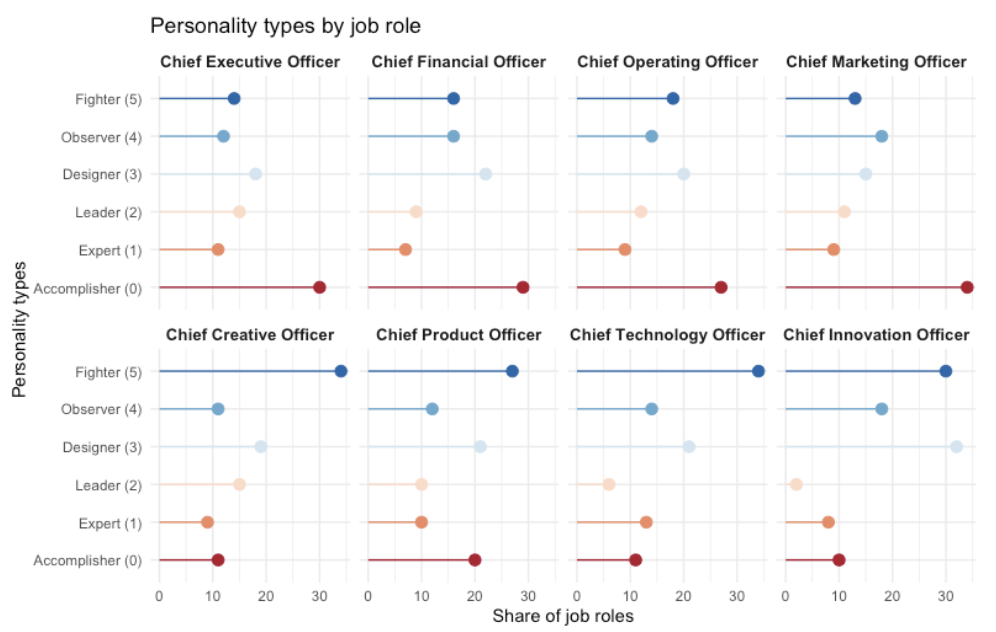}
	\vskip2em
	\caption{
		\textbf{$|$ Occupations within Startups} 
		While Accomplishers are often CEOs, CFOs or COOs, Fighters tend to be CTOs, CPOs, CCOs. 		
	}
	\label{fig:ExtendedFig6}
\end{figure}

\subsection*{Occupations within startups}
Information about personality traits not only helps to distinguish between individuals who tend to be founders of startup companies and employees, but it also correlates with the job role that founders will take in the startup companies they establish. For example, Extended Data Figure~\ref{fig:ExtendedFig6} shows the distribution of the six founder personality clusters by eight typical job roles in startup companies. 

Two personality types are most clearly related to particular job roles. Accomplishers (\#0) cluster in the roles of Chief Executive Officer, Chief Financial Officer, Chief Operating Officer, and Chief Marketing Officer. In contrast, Fighters (\#5) are most prevalent in the roles of Chief Creative Officer, Chief Product Officer and Chief Technology Officer. 

\newpage
\begin{figure}[htbp]
	\centering
	\includegraphics[width=0.76\linewidth]{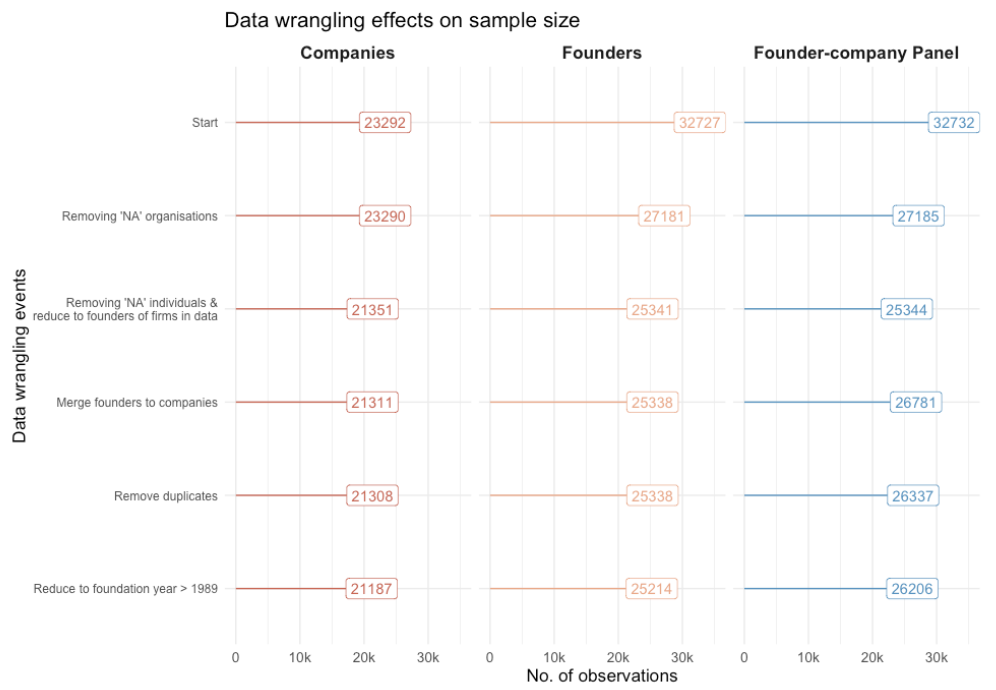}
	\vskip2em
	\caption{
		\textbf{$|$ Data Wrangling for Multifactor Analysis} 
		Effects of the data cleaning on the sample size. Five preparatory steps reduce the data set to 25,214 founders with inferred personality traits who have been involved in founding 21,187 startup companies.}
	\label{fig:ExtendedFig7}
\end{figure}

To use the founder data from Crunchbase described above (32k profiles) for the success prediction, we needed to conduct several preparatory steps, which led to a reduction of the final number of observations as outlined in Extended Data Figure~\ref{fig:ExtendedFig7}. 

The aim is to create a company-founder panel from the Crunchbase data based on exact founder names and company URLs as identifiers. Starting with 32,727 profiles corresponding to 23,292 companies, we removed organisations without names, reducing the data set to 27,181 founders and 23,290 companies. As a next step, we kept only those founders in the data set, founding the 23,290 companies in the data (via the ‘founders’ column), yielding a total of 25,341 founders and 21,351 companies. Merging these founders with the companies led to a further reduction of the data set to 25,338 founders and 21,311 companies. The merging also resulted in some duplicates because of the identical names of some founders. These duplicates were removed by keeping only those company-founder combinations where the company of each potentially duplicated founder was mentioned either as their primary organisation or in their biography. This step did not affect the number of founders but reduced the number of companies by three, which could not unambiguously be assigned to any individual. As the last step, we removed companies that were founded before 1990, leading to a final data set of 25,214 founders involved in the foundation of 21,187 companies. 

\newpage
\begin{figure}[htbp]
	\centering
	\includegraphics[width=0.76\linewidth]{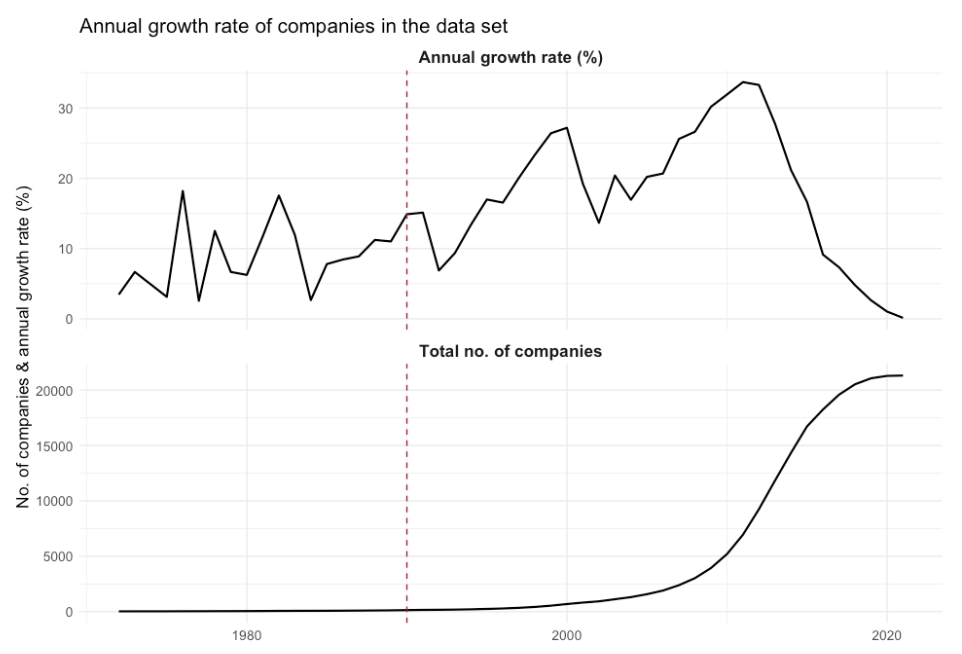}
	\vskip2em
	\caption{
		\textbf{$|$ Foundation Year} 
		Number of companies in the data set by foundation year. Following the approach taken by Bonaventura et al. (2020), we restrict the data set to those companies founded from 1990 onwards.  		
	}
	\label{fig:ExtendedFig8}
\end{figure}

In reducing the data set to those companies that were founded from 1990 (see Extended Data Figure~\ref{fig:ExtendedFig8}) onwards, we aimed at limiting the potential bias that could arise from having companies in the data set that cannot be considered as startup companies because of their age. Therefore, this additional restriction removes less than 0.6\% of the companies in our data set.

\begin{figure}[htbp]
	\centering
	\includegraphics[width=0.76\linewidth]{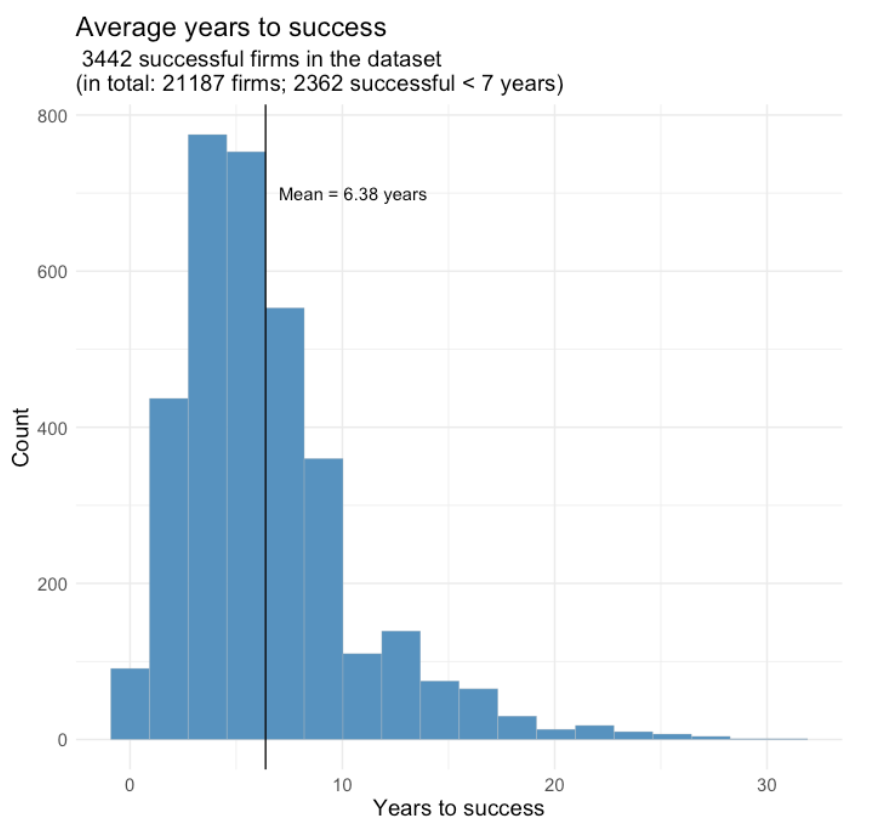}
	\vskip2em
	\caption{
		\textbf{$|$ Years to success} 
		Histogram of the variable ‘average years to success’ based on 3,442 successful firms in the data set. The mean time to success is 6.38 years.  		
	}
	\label{fig:ExtendedFig9}
\end{figure}

In total, 3,442 of 21,187 companies (16\%) in the data set have been successful according to the criterion used by Bonaventura et al.\cite{bonaventura2020predicting}. On average, successful companies needed 6.38 years to become successful (see Extended Data Figure~\ref{fig:ExtendedFig9}). 

\newpage

The final data set is a panel with 26,202 observations, i.e. combinations of 25,214 founders involved in founding 21,187 companies. For each data point, we observe a total of 104 variables. Of those, 15 variables relate to the organisations and cover aspects such as a company identifier, description, industry categories, location, and foundation year. In addition, there are six variables related to success in the data: success date \& type, success indicator variable, years to success since founded, and an indicator if success occurred within the first seven years after foundation. Eight variables refer to the founders: name, Twitter, biography, primary organisation, primary job role, gender, and social media; 75 variables present different characteristics of the inferred personalities: personality type, individual facets, Big 5 traits, etc.

\newpage
\begin{figure}[htbp]
	\centering
	\includegraphics[width=0.76\linewidth]{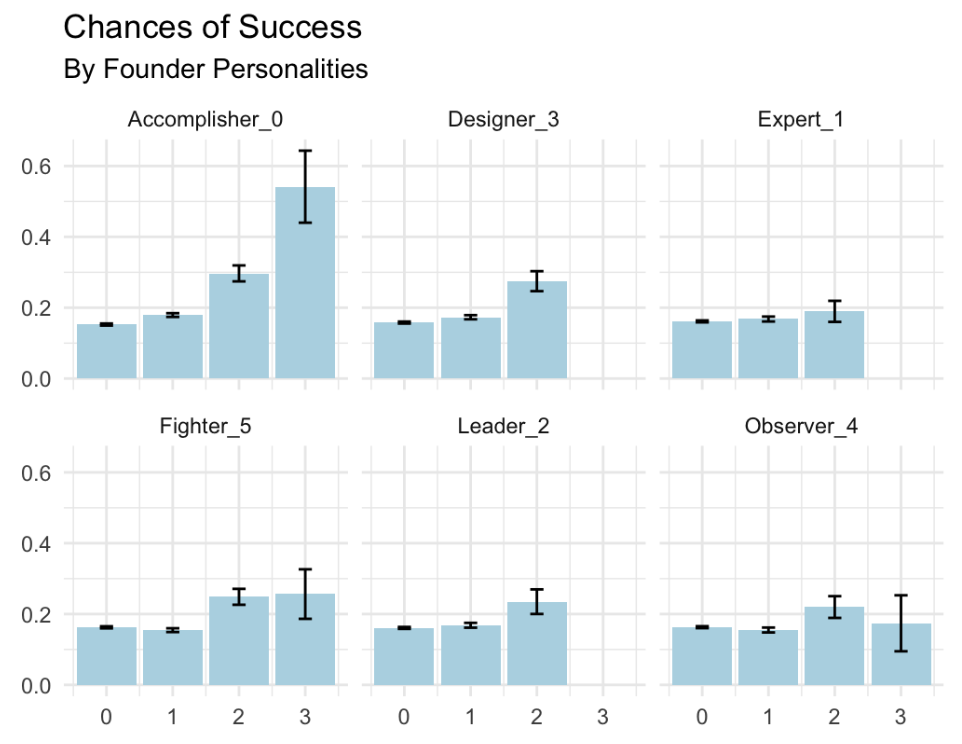}
	\vskip2em
	\caption{
		\textbf{$|$ Founder Team and Success} 
		Firms with multiple founders of the same personality types have higher chances of success.  		
	}
	\label{fig:ExtendedFig14}
\end{figure}

\newpage
\begin{figure}[htbp]
	\centering
	\includegraphics[width=0.76\linewidth]{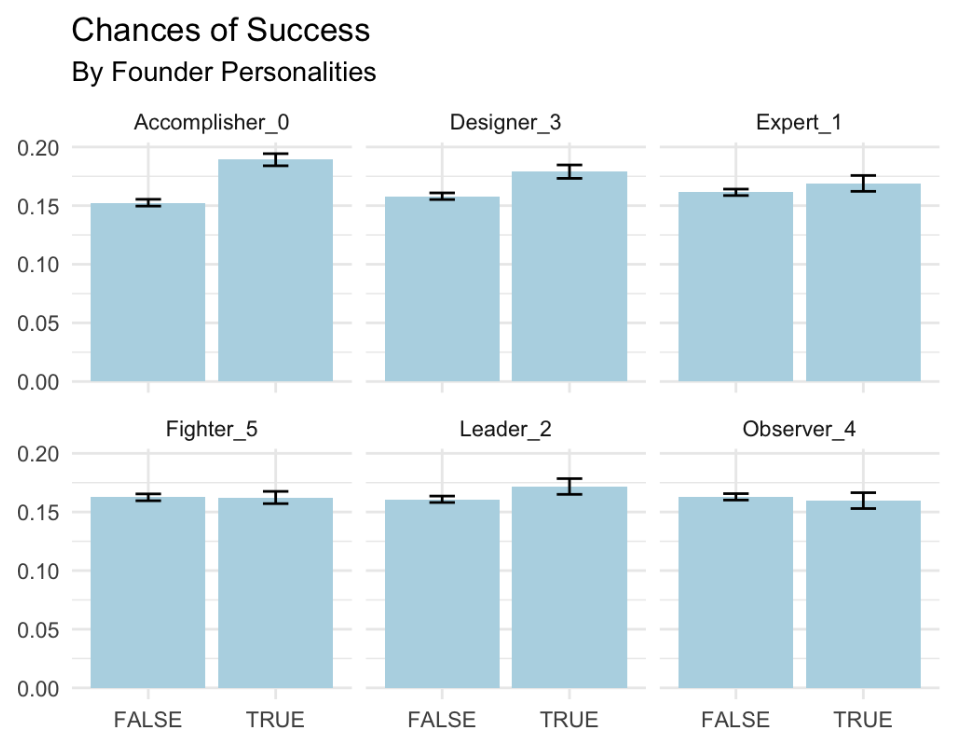}
	\vskip2em
	\caption{
		\textbf{$|$ Founder types and success} 
		Firms with specific founder personalities have higher chances of success - most significant for personalities of the “Accomplisher” type.	
	}
	\label{fig:ExtendedFig15}
\end{figure}

\newpage
\begin{figure}[htbp]
	\centering
	\includegraphics[width=0.76\linewidth]{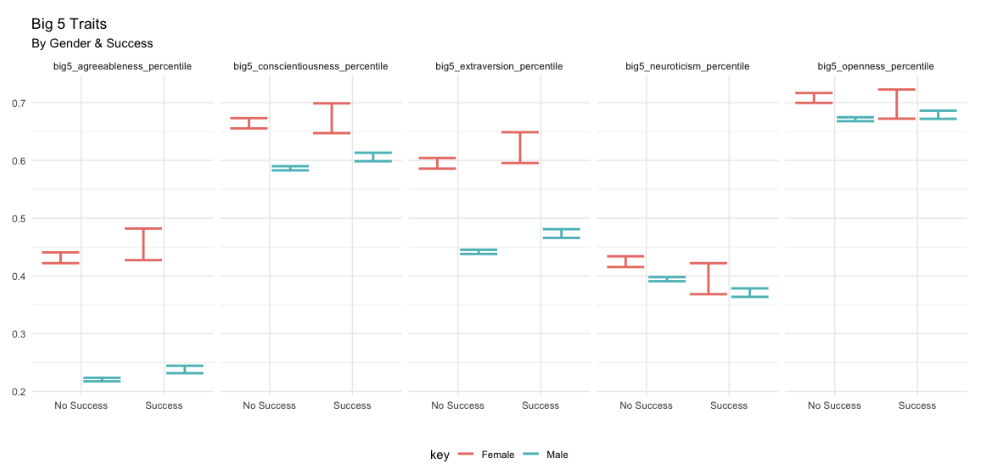}
	\vskip2em
	\caption{
		\textbf{$|$ Gender bias in startups} 
		We noted Successful female founders are more similar to the successful male founder. 	
	}
	\label{fig:ExtendedFig18}
\end{figure}

While the multifactor modelling shows having male founders increases a startup's chance of success, this is likely primarily due to the significant gender bias in venture funding. In our data, female solo-founders received, on average, 20\% less total funds than their male counterparts, while female co-founded teams received less than half the funding on average as all-male teams.  

The amount of venture funds and angel investors specifically targeting female founders has grown rapidly since 2006, such as Women’s Venture Capital Fund (Founded in Portland, Oregon in 2011);  Female Founders Fund (founded in NY in 2014) and Halogen Ventures (founded in 2016 in LA). This will likely address some of these inequities over the current cohort of startups.

\newpage
\begin{figure}[htbp]
	\centering
	\includegraphics[width=0.76\linewidth]{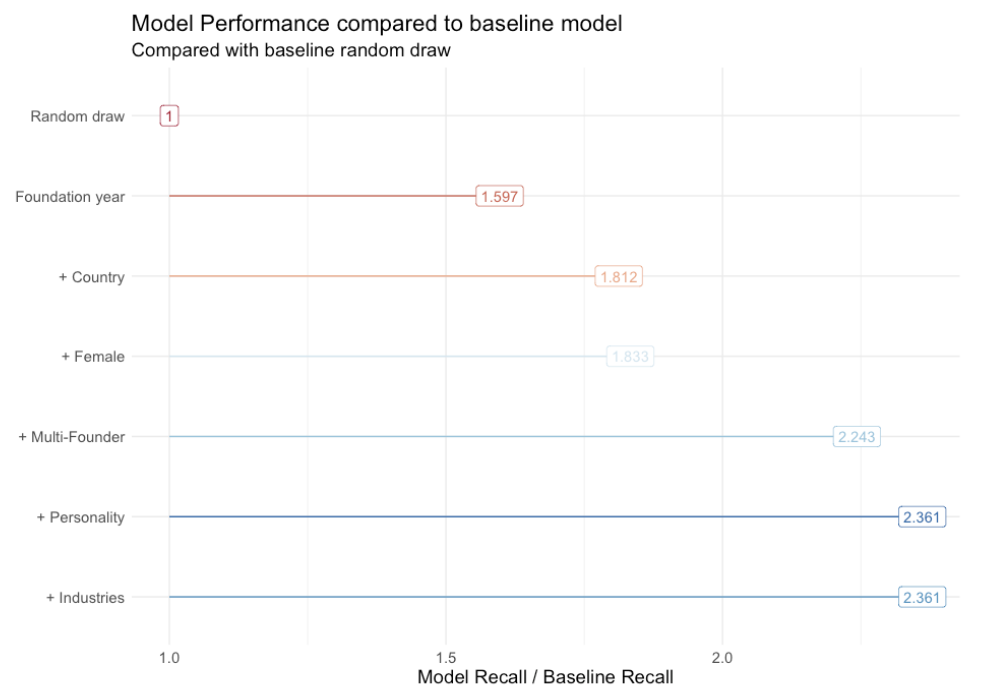}
	\vskip2em
	\caption{
		\textbf{$|$ Recall Performance} 
		Comparison of prediction performance according to Recall metric of different startup success prediction models (GLMs): (0) Null model of random draw, (1) simple model including only the foundation year as a predictive variable, (2) model 1 plus country, (3) model 2 plus female indicator variable, (4) model 3 plus count of a number of founders, (5) model 4 plus personality traits, (6) model 5 plus industries.}
	\label{fig:ExtendedFig19}
\end{figure}

On the one hand, the correlations between external and internal factors and success, on the other hand, are visible when comparing different machine learning models that predict startup success. For example, Extended Data Figure~\ref{fig:ExtendedFig19} shows the predictive performance of six logistic regression models compared to a baseline random draw model. According to the recall Machine Learning Performance metric, the best-performing models (5) and (6) are more than 130\% better than random draw. These models include several explanatory variables: foundation year, country, female indicator variable, the number of founders, as well as personality types (model 5) plus industries (model 6).

\newpage
\begin{figure}[htbp]
	\centering
	\includegraphics[width=0.76\linewidth]{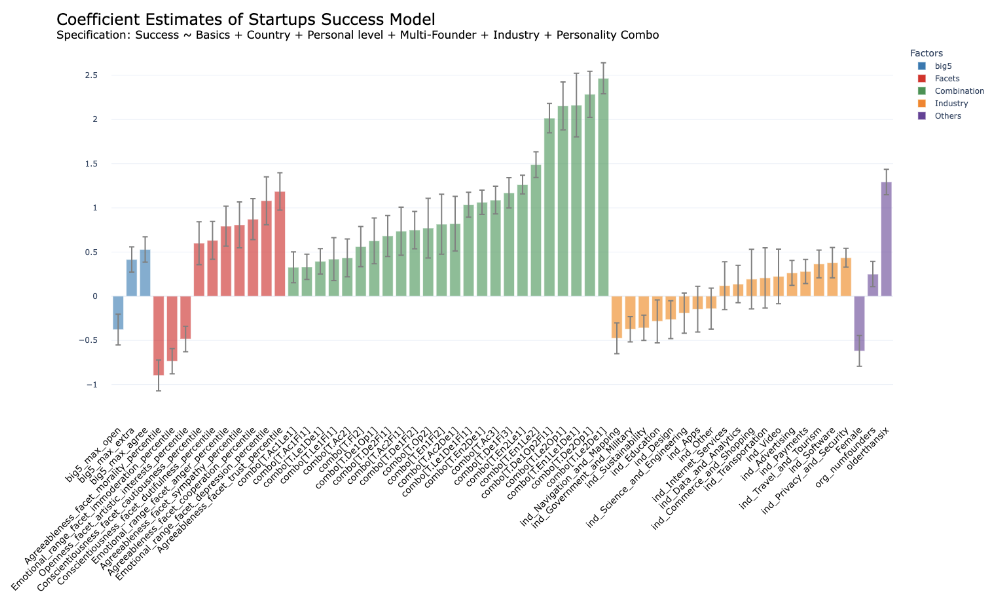}
	\vskip2em
	\caption{
		\textbf{$|$ Ensemble Model} 
		Multifactor modelling reveals the relative significance of different Firm-level, Founder-level, and Founder-Team level features on startup success and illustrates how personality-diverse larger teams have one of the most significant impacts on chances of success.	
	}
	\label{fig:ExtendedFig20}
\end{figure}

\newpage
\begin{figure}
  \includegraphics[width=0.7\textwidth]{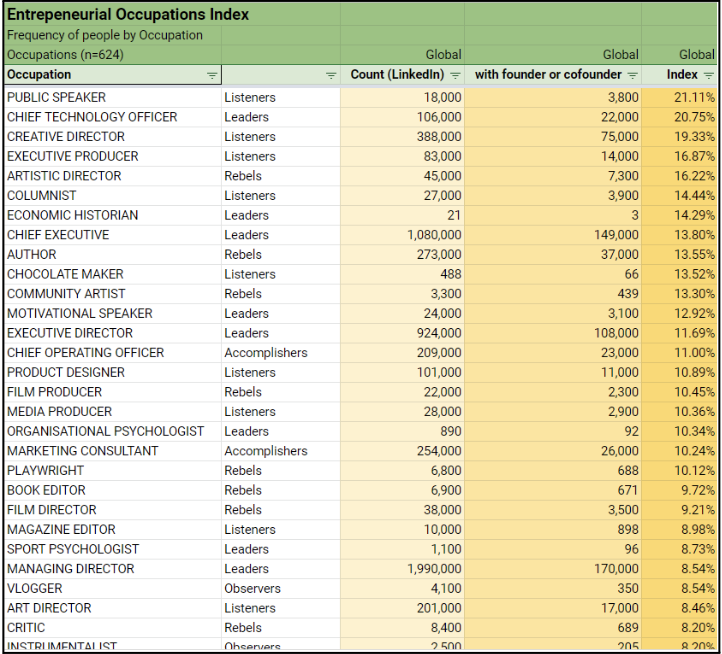}\label{fig:Figure21a}\\[3ex]
  \includegraphics[width=0.7\textwidth]{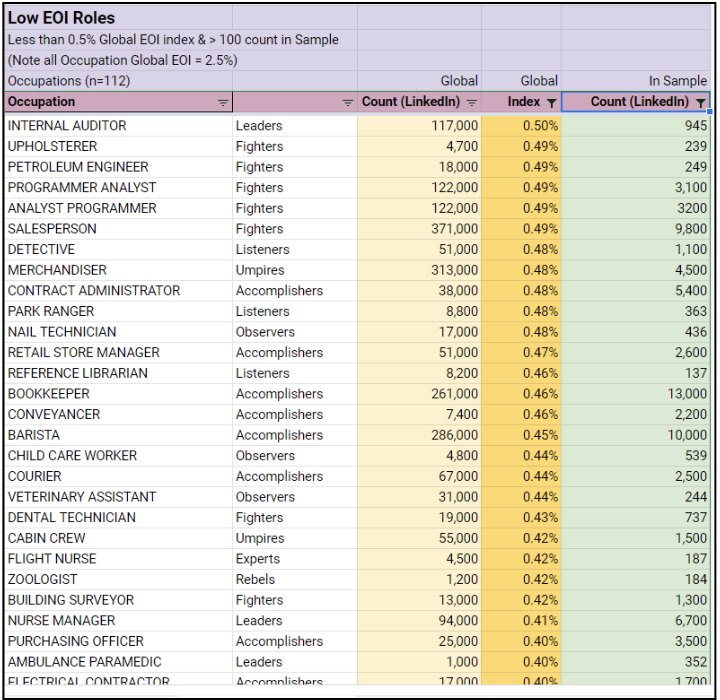}\label{fig:Figure21b}
  \caption{\textbf{$|$ Entrepreneurial Occupations Index} }\label{fig:ExtendedFig21}
\end{figure}

\newpage
\begin{figure}[htbp]
	\centering
	\includegraphics[width=0.76\linewidth]{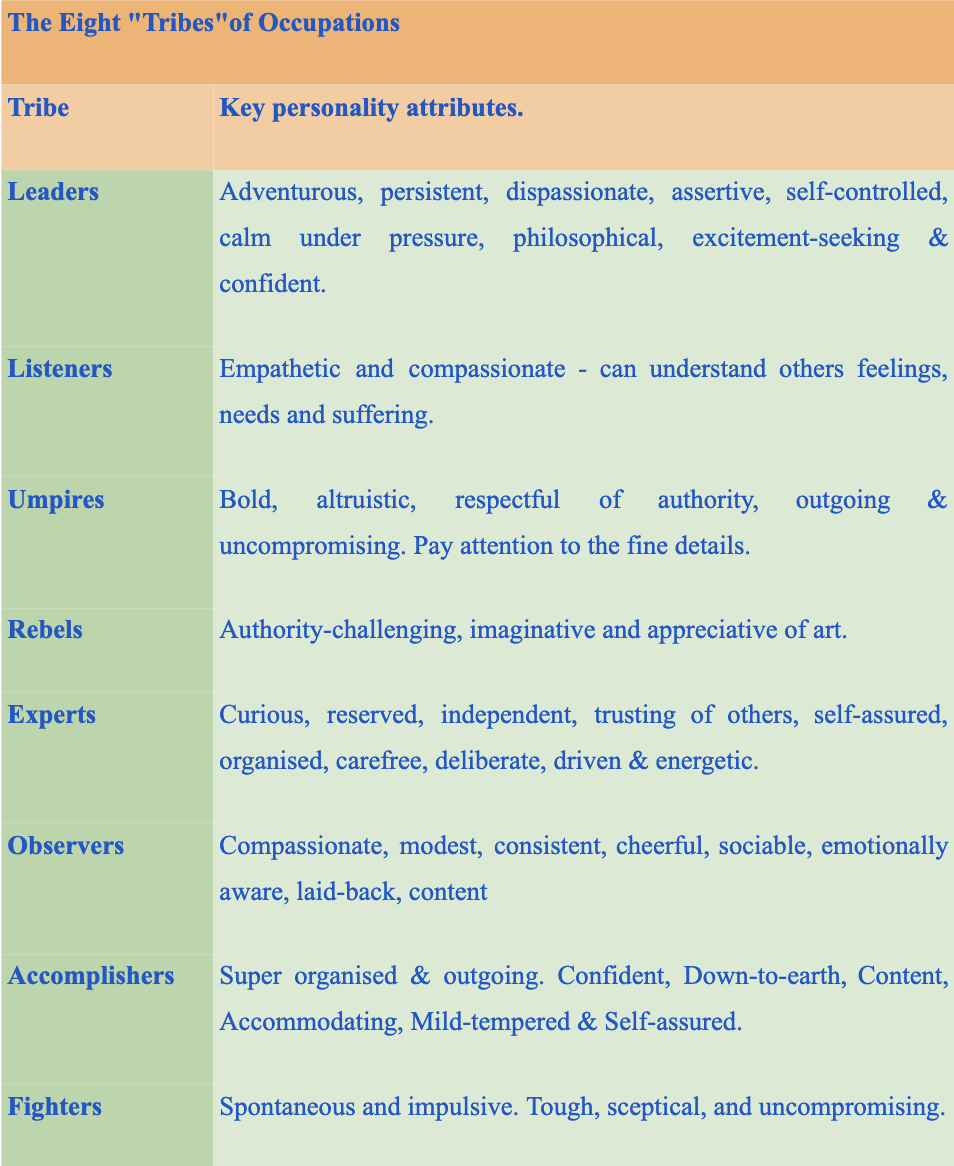}
	\vskip2em
	\caption{
		\textbf{$|$ Eight Tribes\cite{mccarthy2022occupation}}  	
	}
	\label{fig:ExtendedFig22}
\end{figure}

\newpage
\begingroup
\footnotesize
\singlespacing
\begin{longtable}{ C{5cm}  C{3cm}  C{3cm}  C{3cm} }
\caption[$|$ Founder Facet Footprint.]{
    \textbf{$|$ Founder Facet Footprint.}
    Features that distinguish successful founders (n=4.4k) from successful employees (n=6k). Note all 30 Big 5 facets are significant at p $<$ 0.05; however, Artistic Interests and Altruism are not significant at p $<$ 0.01. 
}\\
    
    \toprule
    \makecell{\textbf{Big 5 Personality Facets}} & \makecell{\textbf{Cohen's D}}  & \makecell{\textbf{Effect Size}} & \makecell{\textbf{p Value}}\\
    \hline
    \endhead
   Openness (adventurousness) & 0.92 & Large & 0.00\\
  \hline
   Agreeableness (modesty) & -0.79 & Large & 0.00\\
   \hline
   Extraversion (activity\_level) & 0.78 & Large & 0.00\\
    \hline
   Emotional range (anxiety) & -0.77 & Large & 0.00\\
    \hline
   Emotional range (immoderation) & -0.73 & Large & 0.00\\
  \hline
   Agreeableness (trust) & 0.72 & Large & 0.00\\
   \hline
   Emotional range (anger) & -0.68 & Large & 0.00\\
   \hline
   Emotional range (depression) & -0.68 & Large & 0.00\\
   \hline
   Agreeableness (cooperation) & 0.67 & Large & 0.00\\
   \hline
   Openness (emotionality) & -0.67 & Large & 0.00\\
   \hline
   Emotional range (vulnerability) & -0.63 & Medium & 0.00\\
   \hline
   Conscientiousness (achievement\_striving) & 0.59 & Medium & 0.00\\
   \hline
   Conscientiousness (self\_discipline) & 0.55 & Medium & 0.00\\
   \hline
   Conscientiousness (cautiousness) & 0.50 & Medium & 0.00\\
   \hline
   Openness (intellect) & 0.45 & Medium & 0.00\\
   \hline
   Conscientiousness (self\_efficacy) & 0.43 & Medium & 0.00\\
   \hline
   Extraversion (assertiveness) & 0.43 & Medium & 0.00\\
   \hline
   Openness (liberalism) & 0.43 & Medium & 0.00\\
   \hline
   Agreeableness (sympathy) & -0.43 & Medium & 0.00\\
   \hline
   Conscientiousness (dutifulness) & 0.41 & Medium & 0.00\\
   \hline
   Conscientiousness (orderliness) & 0.31 & Small & 0.00\\
   \hline
   Extraversion (friendliness) & 0.24 & Small & 0.00\\
   \hline
   Extraversion (excitement\_seeking) & -0.21 & Small & 0.00\\
   \hline
   Emotional range (self\_consciousness) & -0.16 & Trivial & 0.00\\
   \hline
   Openness (imagination) & -0.15 & Trivial & 0.00\\
   \hline
   Extraversion (gregariousness) & 0.15 & Trivial & 0.00\\
   \hline
   Agreeableness (morality) & 0.12 & Trivial & 0.00\\
   \hline
   Extraversion (cheerfulness) & -0.08 & Trivial & 0.00\\
   \hline
   Openness (artistic\_interests) & 0.04 & Trivial & 0.00\\
   \hline
   Agreeableness (altruism) & -0.04 & Trivial & 0.01\\
   \bottomrule
   \hline\label{tab:ExtendedTable1}
\end{longtable}
\endgroup

\end{document}